\documentclass[traditabstract,longauthor]{aa} 
\usepackage{natbib}
\usepackage{fixltx2e} 
\usepackage{graphicx}
\usepackage{hyperref}
\usepackage{txfonts}

\begin{document}

\title{A Fast Very High Energy $\gamma$-ray Flare from BL Lacertae during a Period of Multiwavelength activity in June 2015} 



%
\author{
MAGIC Collaboration \and
V.~A.~Acciari\inst{1} \and
S.~Ansoldi\inst{2,21} \and
L.~A.~Antonelli\inst{3} \and
A.~Arbet Engels\inst{4} \and
D.~Baack\inst{5} \and
A.~Babi\'c\inst{6} \and
B.~Banerjee\inst{7} \and
P.~Bangale\inst{8} \and
U.~Barres de Almeida\inst{9} \and
J.~A.~Barrio\inst{10}
% ...
}

\offprints{
  E. Lindfors \email{elilin@utu.fi}, M. Vazquez Acosta \email{monicava@iac.es}, S. Tsujimoto \email{shimpei.tsujimoto@gmail.com}, Filippo D'Ammando \email{dammando@ira.inaf.it}}

\institute { Inst. de Astrof\'isica de Canarias, E-38200 La Laguna, and Universidad de La Laguna, Dpto. Astrof\'isica, E-38206 La Laguna, Tenerife, Spain
\and Universit\`a di Udine, and INFN Trieste, I-33100 Udine, Italy
\and National Institute for Astrophysics (INAF), I-00136 Rome, Italy
\and ETH Zurich, CH-8093 Zurich, Switzerland
\and Technische Universit\"at Dortmund, D-44221 Dortmund, Germany
\and Croatian MAGIC Consortium: University of Rijeka, 51000 Rijeka, University of Split - FESB, 21000 Split, University of Zagreb - FER, 10000 Zagreb, University of Osijek, 31000 Osijek and Rudjer Boskovic Institute, 10000 Zagreb, Croatia.
\and Saha Institute of Nuclear Physics, HBNI, 1/AF Bidhannagar, Salt Lake, Sector-1, Kolkata 700064, India
\and Max-Planck-Institut f\"ur Physik, D-80805 M\"unchen, Germany
\and Centro Brasileiro de Pesquisas F\'isicas (CBPF), 22290-180 URCA, Rio de Janeiro (RJ), Brasil
\and Unidad de Part\'iculas y Cosmolog\'ia (UPARCOS), Universidad Complutense, E-28040 Madrid, Spain
\and University of \L\'od\'z, Department of Astrophysics, PL-90236 \L\'od\'z, Poland
\and Deutsches Elektronen-Synchrotron (DESY), D-15738 Zeuthen, Germany
\and Institut de F\'isica d'Altes Energies (IFAE), The Barcelona Institute of Science and Technology (BIST), E-08193 Bellaterra (Barcelona), Spain
\and Universit\`a di Siena and INFN Pisa, I-53100 Siena, Italy
\and Universit\`a di Pisa, and INFN Pisa, I-56126 Pisa, Italy
\and Universit\`a di Padova and INFN, I-35131 Padova, Italy
\and Universit\"at W\"urzburg, D-97074 W\"urzburg, Germany
\and Finnish MAGIC Consortium: Tuorla Observatory and Finnish Centre of Astronomy with ESO (FINCA), University of Turku, FIN-20014 University of Turku, Finland, Astronomy Division, University of Oulu, FIN-90014 University of Oulu, Finland
\and Departament de F\'isica, and CERES-IEEC, Universitat Aut\'onoma de Barcelona, E-08193 Bellaterra, Spain
\and Universitat de Barcelona, ICCUB, IEEC-UB, E-08028 Barcelona, Spain
\and Japanese MAGIC Consortium: ICRR, The University of Tokyo, 277-8582 Chiba, Japan; Department of Physics, Kyoto University, 606-8502 Kyoto, Japan; Tokai University, 259-1292 Kanagawa, Japan; RIKEN, 351-0198 Saitama, Japan
% ...
}

\date{
  Received DD MM YYYY / Accepted DD MM YYYY
}

\abstract{The mechanisms producing fast variability of the $\gamma$-ray emission in active galactic nuclei are under debate. The MAGIC telescopes detected a fast very high energy (VHE, E$>100$\,GeV) $\gamma$-ray flare from BL Lacertae on 2015 June 15. The flare had a maximum flux of $(1.5\pm 0.3)\times 10^{-10}$ photons cm$^{-2}$ s$^{-1}$ and halving time of $26\pm8$ minutes. The MAGIC observations were triggered by a high state in the optical and high energy (HE, E$>100$\,MeV) $\gamma$-ray bands. In this paper we present the MAGIC VHE $\gamma$-ray data together with multiwavelength data from radio, optical, X-rays, and HE $\gamma$ rays from 2015 May 1 to July 31. Well-sampled multiwavelength data allow us to study the variability in detail and compare it to the other epochs when fast VHE $\gamma$-ray flares have been detected from this source. Interestingly, we find that the behaviour in radio, optical, X-rays and HE $\gamma$-rays is very similar to  two other observed VHE $\gamma$-ray flares. In particular, also during this flare there was an indication of rotation of the optical polarization angle and of activity at the 43\,GHz core. These repeating patterns indicate a connection between the three events. We also test modelling of the spectral energy distribution, based on constraints from the light curves and VLBA observations, with two different geometrical setups of two-zone inverse Compton models. In addition we model the $\gamma$-ray data with the star-jet interaction model. We find that all of the tested emission models are compatible with the fast VHE $\gamma$-ray flare, but all have some tension with the multiwavelength observations.}



\authorrunning{MAGIC Collaboration et al.}
\titlerunning{BL Lac in 2015}
\maketitle

\section{Introduction}
\label{section:Intro}
Blazars are jetted active galactic nuclei (AGN) with the relativistic jets closely aligned to the line of sight of the observer.
They are the most common extragalactic sources in very high energy (VHE: E $>$ 100\,GeV) $\gamma$ rays\footnote{http://tevcat.uchicago.edu}.

Blazars show two broad peaks in the spectral energy
distributions (SEDs).  The lower energy spectral peak in the optical to X-ray range is commonly associated to synchrotron emission from
relativistic electrons. The spectral
peak in the high-energy (HE: 100\,GeV $>$E $>$ 100\,MeV) to VHE $\gamma$-ray range is widely believed to be produced by inverse Compton (IC) scattering off the synchrotron photons (Synchrotron Self Compton, SSC)
  \citep[see e.g.][]{maraschi92}, and/or IC
scattering with photons from outside the jet in the
external Compton (EC) scenario \citep{dermer,sikora94}.
Hadronic models \citep[e.g.][]{mannheim,mucke}, where a significant role is
played by relativistic protons in the jet, can not be ruled out at the present state of observations.

Based on their optical spectra, blazars are divided into two classes:
flat spectrum radio quasars (FSRQs) that show broad optical emission lines, and
BL Lacertae objects (BL Lacs) characterised by the weakness or even
absence of such emission lines \citep{Weymann91,Stickel91}. In \citet{ghisellini11} a more physical classification scheme between FSRQs and BL Lacs was suggested, based on the luminosity of the broad-line region (BLR) measured in Eddington units.
Depending on the peak frequency of the low-energy bump of the SED, BL Lacs are further subdivided into high- (HBL),
intermediate- (IBL), and low- (LBL) frequency-peaking BL Lac objects, with log $\nu_{\mathrm {peak}}<14$ defining a LBL, $14<$log $\nu_{\mathrm {peak}}<15$ an IBL and log $\nu_{\mathrm {peak}}>15$ for a HBL \citep{PadovaniGiommi95,abdo10}.

BL Lacertae (hereafter BL Lac) is a prototype of the BL Lac objects{\bf,} with a redshift of
z=0.069 \citep{Miller78}, and according to its synchrotron peak frequency is classified as a LBL \citep{Nilsson18} or IBL \citep{ackermann11}. \citet{hervet} recently suggested a classification based on kinematic features of the radio jets, quasi-stationary or knots, and in their work they classified BL Lac as an intermediate source. BL Lac is well
known for its prominent variability in a wide energy range,
in particular in optical and radio bands and has been a target of many multiwavelength campaigns \citep[e.g.][]{Hagen-Thorn02,marscher08,raiteri09,abdo11,raiteri13,Wehrle16}. It shows complicated long-term behaviour and has been suggested to show quasi-periodic variability in optical and HE $\gamma$-ray bands \citep{Sandrinelli}.

The first detection of VHE $\gamma$ rays from BL Lac was reported by the Crimean
Observatory with 7.2\,$\sigma$ significance above 1\,TeV in 1998
\citep{Neshpor01}.  In the meantime, HEGRA observed it in the same
period and obtained only an upper limit \citep{Kranich03}.  Subsequently,
the MAGIC collaboration observed BL Lac for 22.2\,h in 2005 and
for 26 h in 2006, and a VHE $\gamma$ ray signal was discovered in the 2005 data
with an integral flux of 3\% of the Crab Nebula flux above 200 GeV
\citep{Albert07}.  On 2011 June 28, a very rapid TeV $\gamma$-ray flare
from BL Lac was detected by VERITAS.  The flaring activity was
observed during a 34.6 minute exposure, when the integral flux above
200 GeV reached $(3.4\pm0.6)\times 10^{-6}$\,photons m$^{-2}$
s$^{-1}$, roughly 125\% of the Crab Nebula flux \citep{Arlen13}. After that, two more fast VHE $\gamma$-ray flares from BL Lac have been observed, one by MAGIC and one by VERITAS \citep{ATel, ATel_veritas}.

In past years, fast VHE $\gamma$-ray flares have been detected also
from many other AGN: from HBLs \citep{2155,mrk501},
from FSRQs \citep{al11,zacharias} and from
radio galaxies \citep{m87,ic310}. Of the IBLs and LBLs
observed in the VHE $\gamma$ rays, BL Lac is the only one where
sub-hour variability has been detected.  This sub-hour, even minute
scale variability challenges standard models of blazar variability. Many models have been suggested ranging from VHE $\gamma$ rays originating
close to the black hole magnetosphere \citep{ic310,hirotanipu}, to mini-jets from magnetic reconnection \citep[e.g.][]{giannios09,morris18} or star-jet interactions \citep[e.g.][]{ba10} to more
traditional high Doppler factor small blobs \citep{begelman}
travelling in the jet and possibly interacting with larger emission
regions \citep{tavecchiobecerra}.

In this paper we report a detection of a fast flare in VHE
$\gamma$ rays from BL Lac on 2015 June 15 and the
quasi-simultaneous multiwavelength observations. Part of these data were
already presented by \citet{icrc_bllac}. Here we report on
complete results from this observational campaign, compare them with the other two VHE $\gamma$-ray flares detected from the source and discuss
theoretical models that can reproduce the fast variability.

\begin{figure*}
  \centering
  \includegraphics[width=8cm]{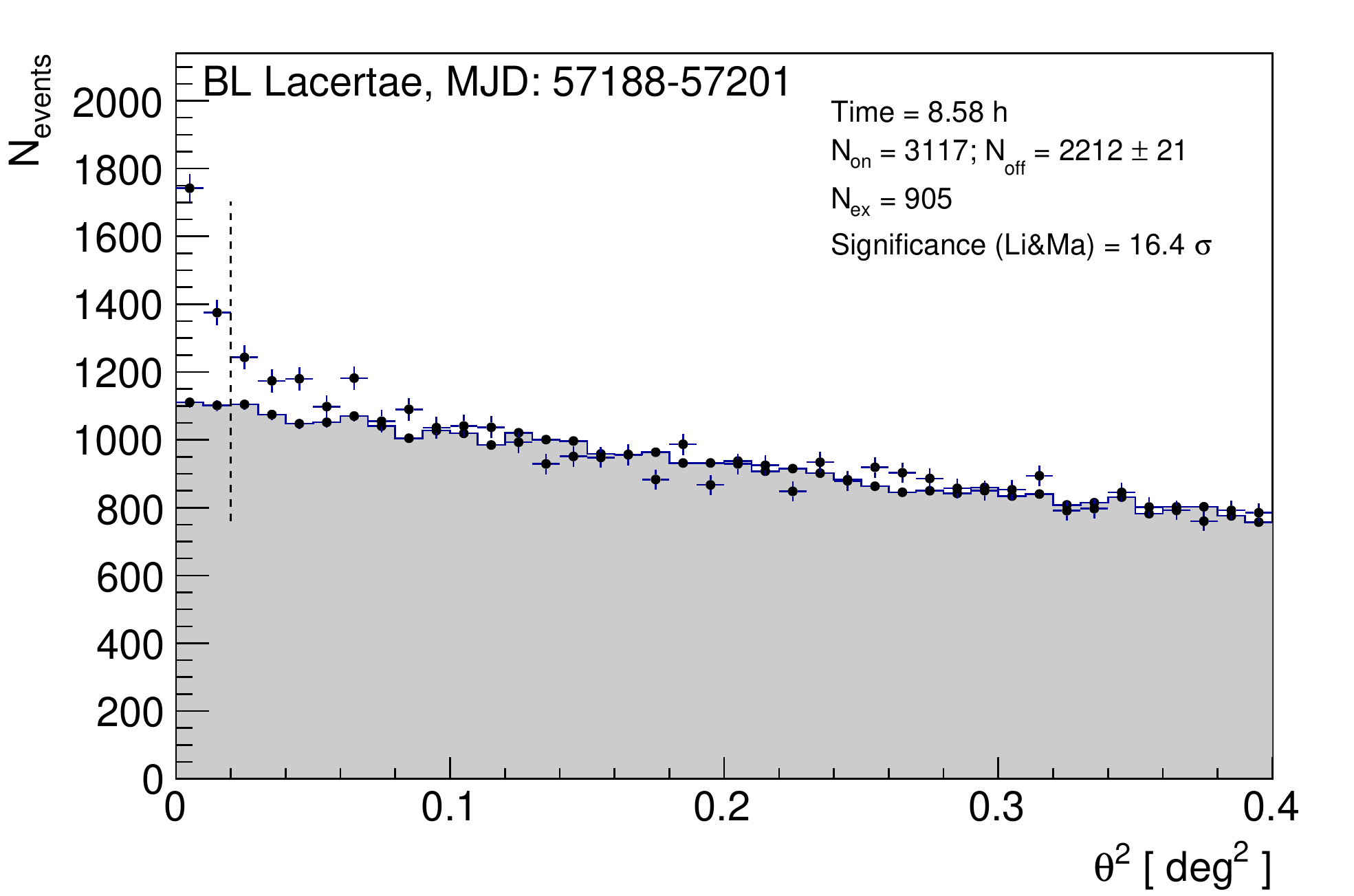}
  \includegraphics[width=8cm]{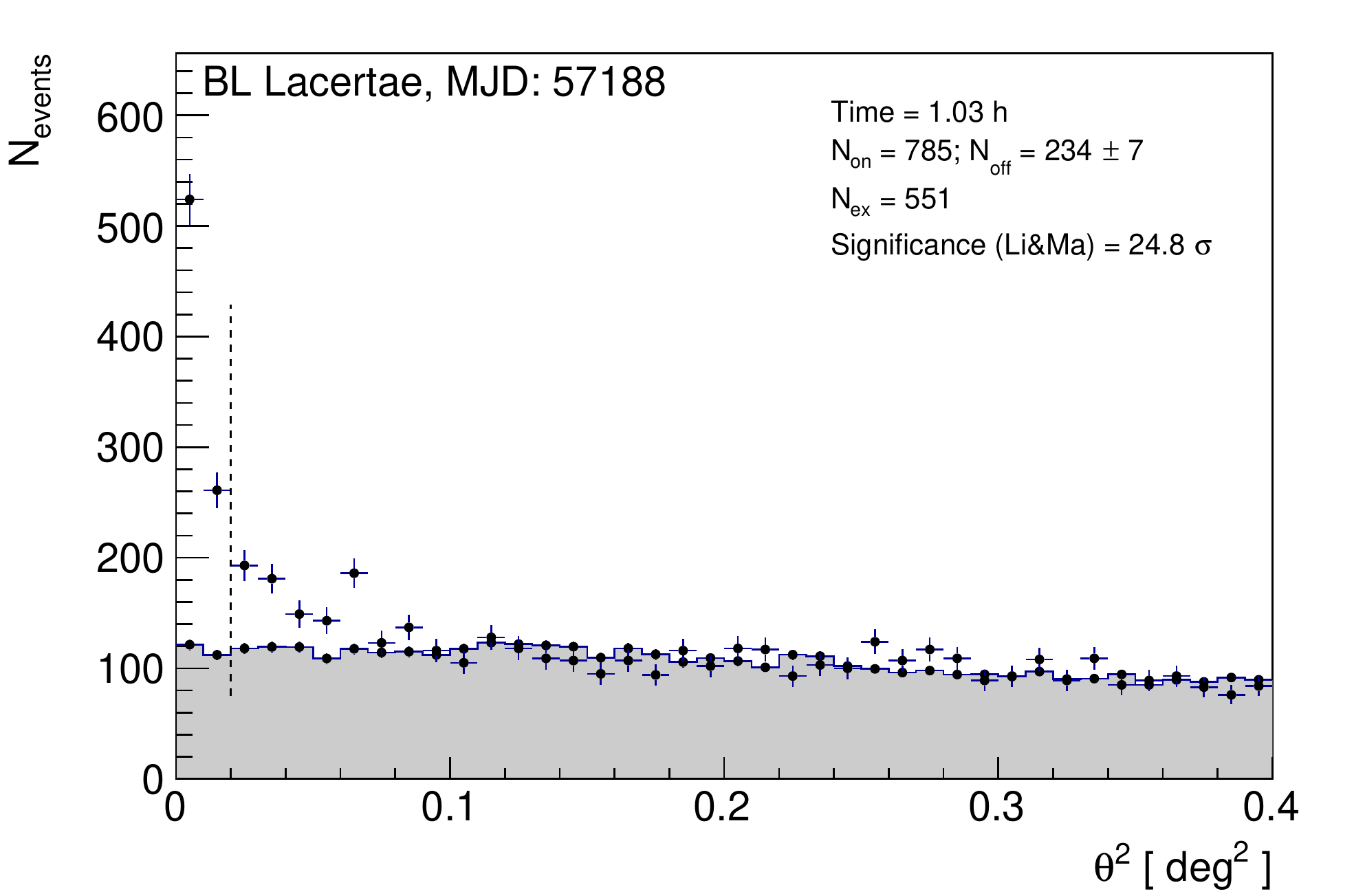}
     \caption{Theta-squared distribution of data taken between 2015 June 15 and 28 (left) and on 2015 June 15 (right). Excess events are shown with filled circles, highlighted with blue crosses, and the normalized off-source events are shown with grey histogram. The energy threshold is $\sim$70\,GeV. The vertical dashed line marks the limit of the signal region at $\theta^2=0.02^\circ$.}
  \label{Theta2}
\end{figure*}

\begin{figure}
  \centering
  \includegraphics[width=0.43\textwidth]{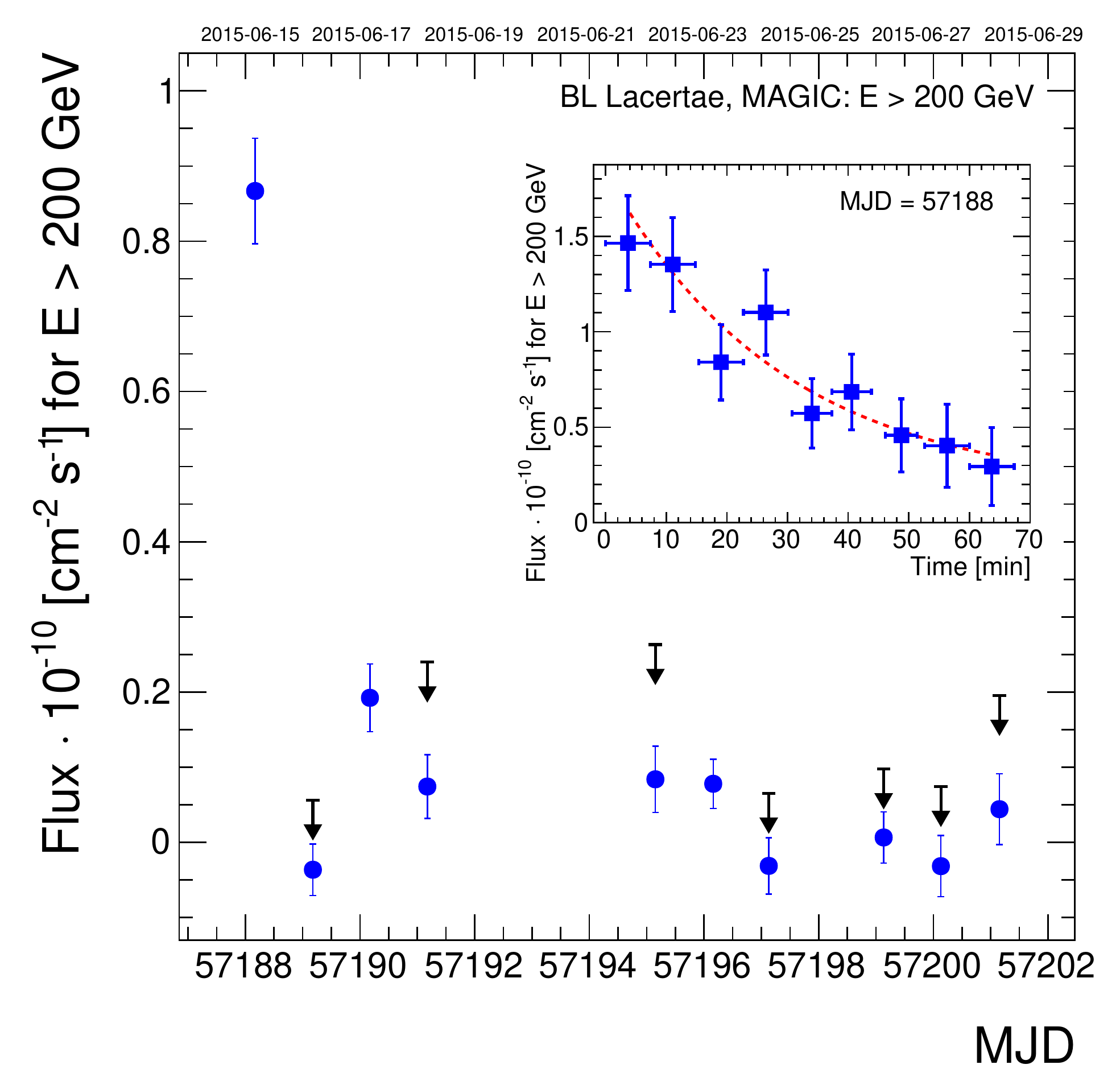}
  \caption{Daily light curve of the VHE $\gamma$-ray emission from BL Lac above 200\,GeV between 2015 June 15 and 28. The time evolution of the flare on MJD 57188 is shown in the inset. Arrows on this figure show the 95\% confidence level upper limits.}
  \label{MAGIC_LC}
\end{figure}

\section{Multiwavelength observations}
\label{section:MWLobs}
\subsection{MAGIC telescopes}
The Major Atmospheric Gamma-ray Imaging Chenrenkov (MAGIC) is a system of two 17\,m Imaging Atmospheric Cherenkov Telescopes (IACTs) located in the Canary Island of La Palma, Spain, at ~2200\,m above sea level.
The low-energy threshold of the MAGIC telescopes (the standard trigger threshold is $\sim$50\,GeV) is an advantage to carry out blazar observations with VHE $\gamma$ rays.
The integral sensitivity for point-like sources with Crab Nebula-like
 spectra above 218\,GeV is (0.66 $\pm$ 0.03)\% of the Crab Nebula flux in 50\,h of observations \citep{Aleksic16}.

Observations of BL Lac 
(R.A. 22h02m43.3s, Dec. +42$^\circ$16'40''; J2000.0) 
were performed during 10 nights between 2015 June 15 and June 28 for a total of 8.58\,h. The observations were triggered by a high state in HE $\gamma$-ray ($F_{E>100\,MeV} >0.5\times10^{-6}$ photons cm$^{-2}$ s$^{-1}$) and optical bands ($F_{R}>20$\,mJy).
Data were taken with zenith angles in the range 14$^\circ$ to 32$^\circ$ that guarantees the lowest energy threshold.
Observations were carried out in the so called wobble mode \citep{Fomin1994}, 
where the telescopes alternated four sky positions every 20 minutes with an offset of 0.4$^\circ$ from the source direction.

The data analysis was performed using the MAGIC analysis and reconstructions software \texttt{MARS} \citep{zanin,Aleksic16}, and following the standard MAGIC analysis chain. The systematic errors are 30\% on flux estimation and $\pm0.15$ on spectral index \citep{Aleksic16}.

\subsection{{\it Fermi}-LAT Data}
\label{FermiData}

The Large Area Telescope (LAT) on board the {\it Fermi Gamma-ray Space Telescope} is a pair-conversion detector
operating from 20 MeV to $>$ 300 GeV. Further details about the {\em Fermi}-LAT are given by \citet{atwood09}. BL Lac is included in all LAT catalogues and also the LAT hard-source catalogs 2FHL (above 50 GeV, \citet{Ackermann16}) and 3FHL (above 10 GeV, \citet{Ajello17}).

The LAT data used in this paper were collected from 2015 May 1 (MJD 57143) to July 31 (MJD 57234). During this time, the LAT instrument operated almost entirely in survey mode. The Pass 8 data \citep{atwood13}, based on a complete and improved revision of the entire LAT event-level analysis, were used. The analysis was performed with the \texttt{ScienceTools} software package version v10r0p5. Only events belonging to the `Source' class (\texttt{evclass=128}, \texttt{evtype=3}) were used. We selected only events within a maximum zenith angle of 90$^\circ$ to reduce contamination from the Earth limb $\gamma$ rays, which are produced by cosmic rays interacting with the upper atmosphere. The LAT analysis was performed with the instrument response functions \texttt{P8R2\_SOURCE\_V6} using a binned maximum-likelihood method implemented in the Science tool \texttt{gtlike}. Isotropic (`iso\_source\_v06.txt') and Galactic diffuse emission (`gll\_iem\_v06.fit') components were used to model the background \citep{acero16}\footnote{http://fermi.gsfc.nasa.gov/ssc/data/access/lat/\\BackgroundModels.html}. The normalization of both components was allowed to vary freely during the spectral fitting.

We analysed a region of interest of $30^{\circ}$ radius centred at the location of BL Lac. We evaluated the significance of the $\gamma$-ray
signal from the source by means of a maximum-likelihood test statistic (TS) defined as TS = 2$\times$(log$L_1$ - log$L_0$), where $L$ is the likelihood of the data given the model with ($L_1$) or without ($L_0$) a point source at the position of BL Lac \citep[e.g.,][]{mattox96}. The source model used in \texttt{gtlike} includes all the point sources from the 3FGL catalogue \citep{3FGL} that fall within $40^{\circ}$ of BL Lac. The spectra of these sources were parametrized by a power-law (PL), a log-parabola (LP), or a super exponential cut-off, as in the 3FGL catalogue. 

A first maximum likelihood analysis was performed over the whole period to remove from the model the sources having TS $< 10$. A second maximum likelihood analysis was performed with the updated source model. In the fitting procedure, the normalization factors and the spectral shape parameters of the sources lying within 10$^{\circ}$ of BL Lac were left as free parameters. For the sources located between 10$^{\circ}$ and 40$^{\circ}$ from our target, we kept the normalization and the spectral shape parameters fixed to the values from the 3FGL catalogue.

\subsection{Swift Satellite}

{\it The Neil Gehrels Swift Observatory}
\citep{gehrels04} carried out 31 observations of BL Lac between 2015 May 2 (MJD 57144) and July 29 (MJD 57232). The observations were performed with all three instruments on board: the X-ray Telescope \citep[XRT;][0.2--10.0 keV]{burrows05}, the Ultraviolet/Optical Telescope \citep[UVOT;][170--600 nm]{roming05} and the Burst Alert Telescope \citep[BAT;][15--150 keV]{barthelmy05}.

The hard X-ray flux of this source turned out to be below the sensitivity of the BAT instrument for such short exposures and therefore the data from this instrument are not included in this work.

\begin{table*}[th]
\centering
\begin{tabular}{lcccc}
\hline
\hline
Date&Observing time& Significance &F($>$200\,GeV)& Upper limit 95\% C.L\\
MJD&h&$\sigma$&10$^{-11}$ photons cm$^{-2}$ s$^{-1}$&10$^{-11}$ photons cm$^{-2}$ s$^{-1}$\\
\hline
57188.2&1.03&24.8& $8.7\pm0.7$&\\
57189.2&0.73&1.8& $-0.4\pm0.3$&0.5\\
57190.2&0.97&5.7& $1.9\pm0.4$&\\
57191.2&0.82&3.3& $0.7\pm0.4$&2.4\\
57195.2&0.65&-0.4& $0.8\pm0.4$&2.6\\
57196.2&1.53&5.1& $0.8\pm0.3$&\\
57197.1&0.65&2.4& $-0.3\pm0.4$&0.7\\
57199.1&0.98&1.3& $0.1\pm0.3$&1.0\\
57200.1&0.56&-1.3&$-0.3\pm0.4$&0.7\\
57201.2&0.71&0.7& $0.4\pm0.5$&2.0\\
\hline
\end{tabular}
\caption{Daily results of the MAGIC observations, including observation times, detection significances, fluxes ($>$200\,GeV) and, in the case of non-significant detection, upper limits with a 95\% C.L.}
\label{MAGIC_results}
\end{table*}

\subsubsection{{\em Swift}-XRT data}

The raw data of these observations were downloaded from the publicly
available SWIFTXRLOG (\textit{Swift}-XRT Instrument
Log)\footnote{\url{https://heasarc.gsfc.nasa.gov/W3Browse/swift/swiftxrlog.html}}. All the observations had been performed in photon counting (PC) mode. Out of the 31 observations, 25 were qualified for further analysis and the data were processed using the procedure described by \citet{2017A&A...608A..68F}, assuming a fixed equivalent Galactic hydrogen column density of $n_{H} = 3.44 \times 10^{21}$ cm$^{2}$ \citep{raiteri09}.

\subsubsection{{\em Swift}-UVOT data}

During the {\em Swift} pointings, the UVOT instrument observed BL Lac in all its optical ($v$, $b$ and $u$) and UV ($w1$, $m2$ and $w2$) photometric bands \citep{poole08,breeveld10}. We analysed the data using the \texttt{uvotsource} task included in the \texttt{HEAsoft} package
(v6.18). Source counts were extracted from a circular region of 5 arcsec radius centred on the source, while background counts were derived from a
circular region of 10 arcsec radius in a nearby source-free region.
Following \citet{raiteri13}, we assumed a flux density of 2.89,
1.30, 0.36, 0.026, 0.020, and 0.017 mJy for the host galaxy in the v, b, u, uvw1, uvm2, uvw2 bands. By considering the source extraction radius used for the UVOT photometry, the host galaxy contribution is about 50 per cent of the {\bf total galaxy} flux, and it is removed from the magnitude values. Note that the host galaxy contribution in UV is negligible with respect to the source flux, even in the low states. The UVOT flux densities are corrected for dust extinction using the E(B--V) value of 0.291 from \citet{schlafly11} and the extinction laws from \citet{cardelli89}.

\subsection{Optical data}
Optical R-band observations were performed as part of the Tuorla blazar monitoring program \footnote{\url{http://users.utu.fi/kani/1m}}. The observations were performed using a 35\,cm Celestron telescope that is attached to the 60\,cm KVA (Kungliga Vetenskapsakademi) Telescope, located at La Palma. The data analysis was performed using standard procedures with a semi-automatic pipeline \citep{Lindfors16,Nilsson18}. The fluxes were corrected for Galactic reddening and the host galaxy contribution was subtracted using values from \citet{schlafly11, Nilsson07}.

Optical polarization observations were performed with the Nordic Optical Telescope (NOT) at La Palma, Steward Observatory, Perkins, AZT-8+ST7 and Calar Alto 2.2\,m Telescopes. The NOT observations were performed as part of the dedicated observing program to support MAGIC blazar observations. The observations and data analysis were done as in \cite{Hovatta16}. The Steward Observatory data are publicly available; the observations and data analysis methods are described in \citet{smith09}. Perkins and AZT-8+ST7 data were analyzed as in \citet{Jorstad13,Larionov08}. Calar Alto data were acquired as part of the MAPCAT project{\footnote{\url{http://www.iaa.es/~iagudo/_iagudo/MAPCAT.html}}}, see \citet{Agudo12}.

\subsection{Radio data}
BL Lac is part of many radio monitoring programs and in this paper data from the Owens Valley Radio Observatory (OVRO), Mets\"ahovi and Boston blazar monitoring programs are included.

OVRO blazar monitoring program observations are performed with the OVRO 40 meter telescope at 15\,GHz. The observations program and the data analysis are described by \cite{richards11}. The Mets\"ahovi radio telescope is 13.7 meters in diameter and located in Kylm\"al\"a, Finland. The observations are performed at 37\,GHz and data analysis is described by \cite{terasranta98}.

The Boston blazar monitoring program uses the Very Long Baseline Array to perform monthly monitoring of a sample of blazars at 43\,GHz. The observations and data analysis were done as in \cite{jorstad05,jorstad17}. We used a set of calibrated VLBA data at 43\,GHz from the Boston University website \footnote{\url{https://www.bu.edu/blazars/VLBA$_$GLAST/bllac.html}} for eight epochs in 2015 and analyzed both total and polarized intensity images. Note that the polarized intensity images in August and September 2015 have a higher polarized intensity noise level due to poor weather at short baselines.

\begin{figure}
  \centering
  \includegraphics[width=0.43\textwidth]{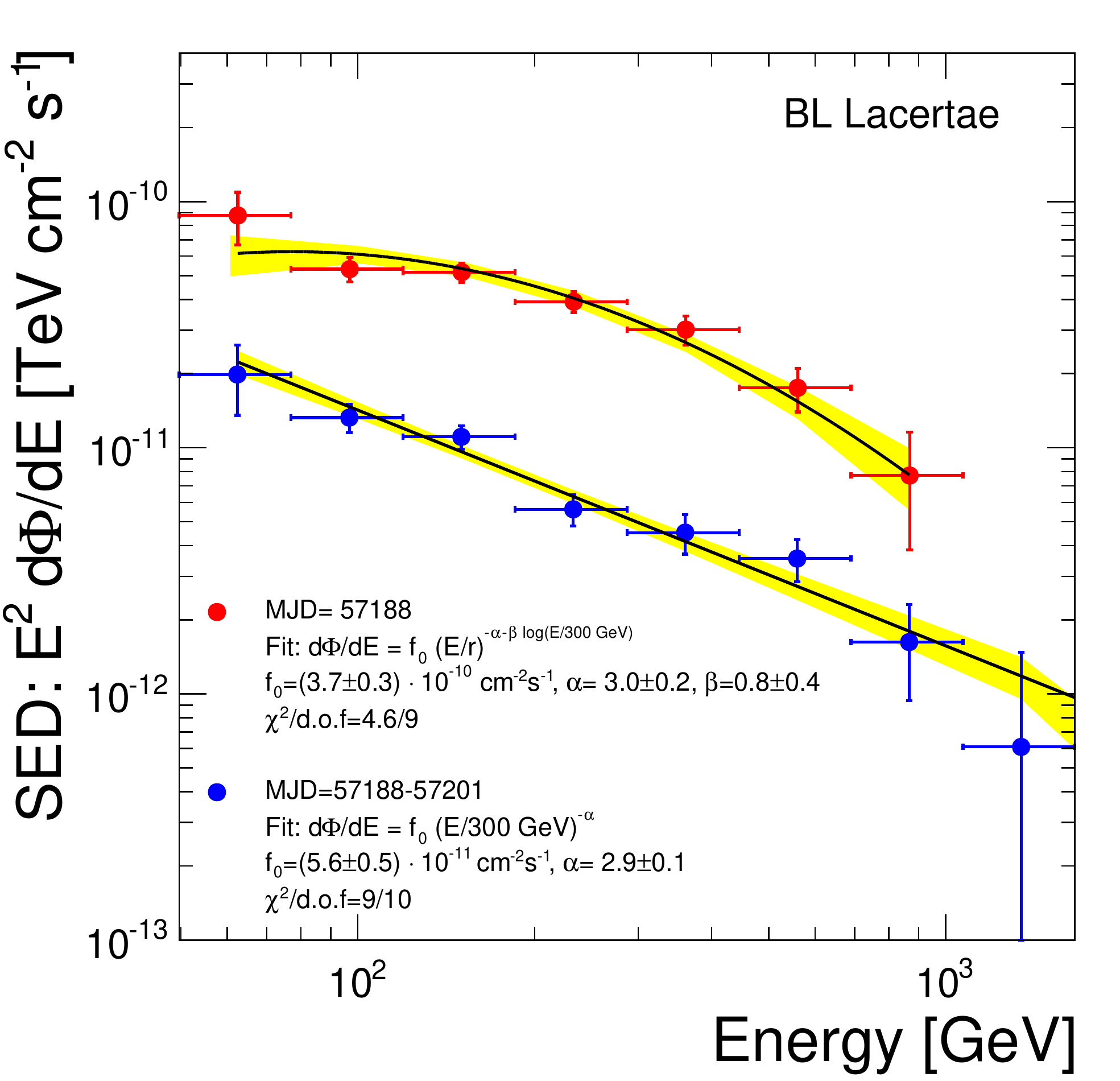}
     \caption{VHE differential energy spectrum of BL Lac for 2015 June between 15 and 28 (blue) and on 2015 June 15 (red). The spectra have been corrected for extragalactic background light absorption using the model of \citet{do11}. The spectra are fitted with PL and LP models and are shown with black solid lines. The yellow band shows the uncertainty of the fit.}
  \label{MAGIC_SED}
\end{figure}

\begin{figure*}
  \centering
  \includegraphics[width=0.68\textwidth]{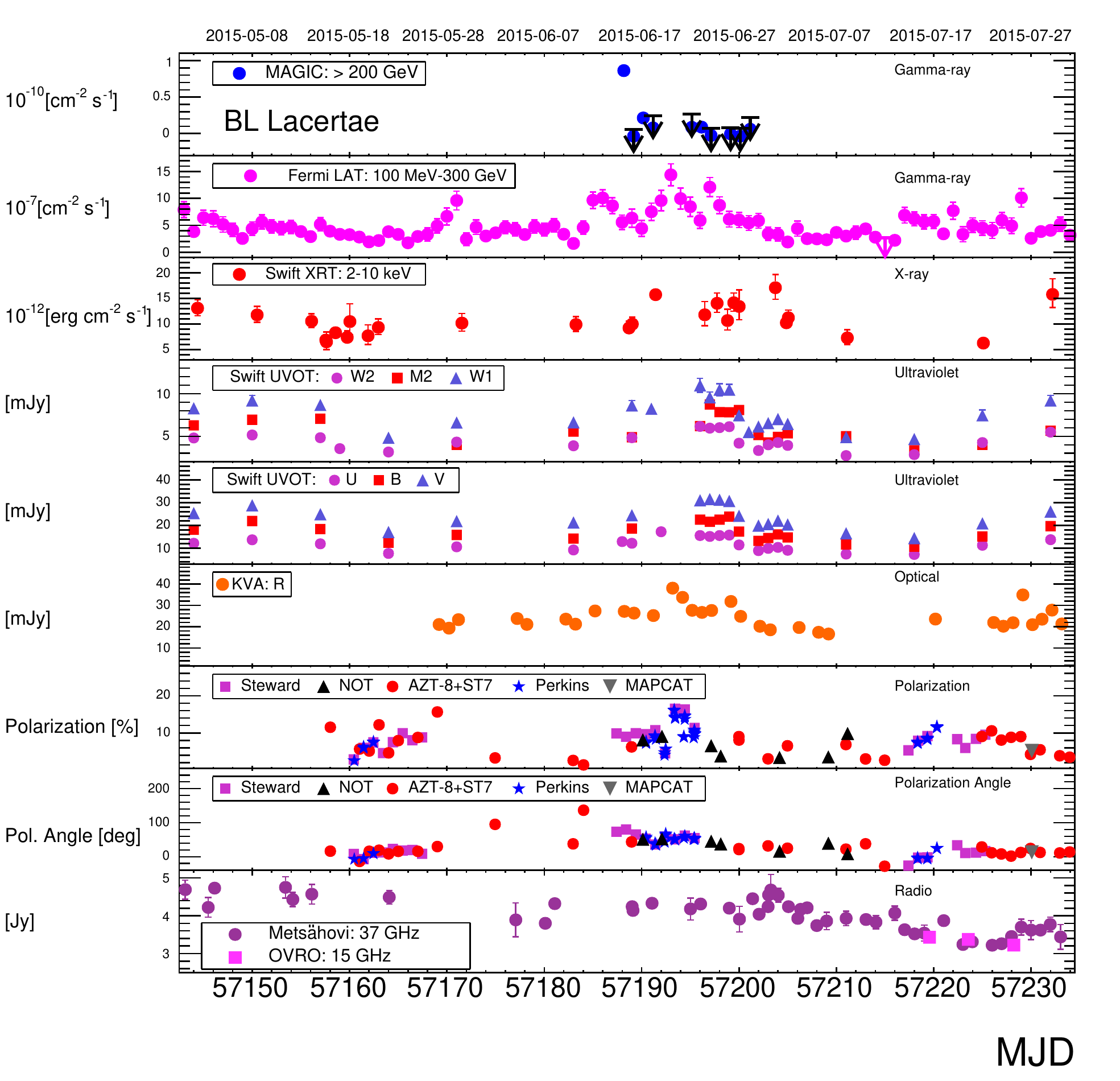}
  \caption{Multiwavelength light curves of BL Lac in the range from MJD 57143 (2015 May 01) to 57234 (2015 July 31). From top to bottom: MAGIC, {\it Fermi}-LAT, {\it Swift}-XRT, UV band of {\it Swift}-UVOT, Optical band of {\it Swift}-UVOT, Optical R-band data from KVA, Polarization data from Steward, NOT, Perkins, AZT-8+ST7 and MAPCAT, Mets\"ahovi and OVRO. MAGIC and {\it Fermi}-LAT data are photon fluxes. Arrows on this figure show the 95\% confidence level upper limits.}
  \label{MWL_LC_long}
\end{figure*}

\begin{figure*}
  \centering
\includegraphics[width=0.68\textwidth]{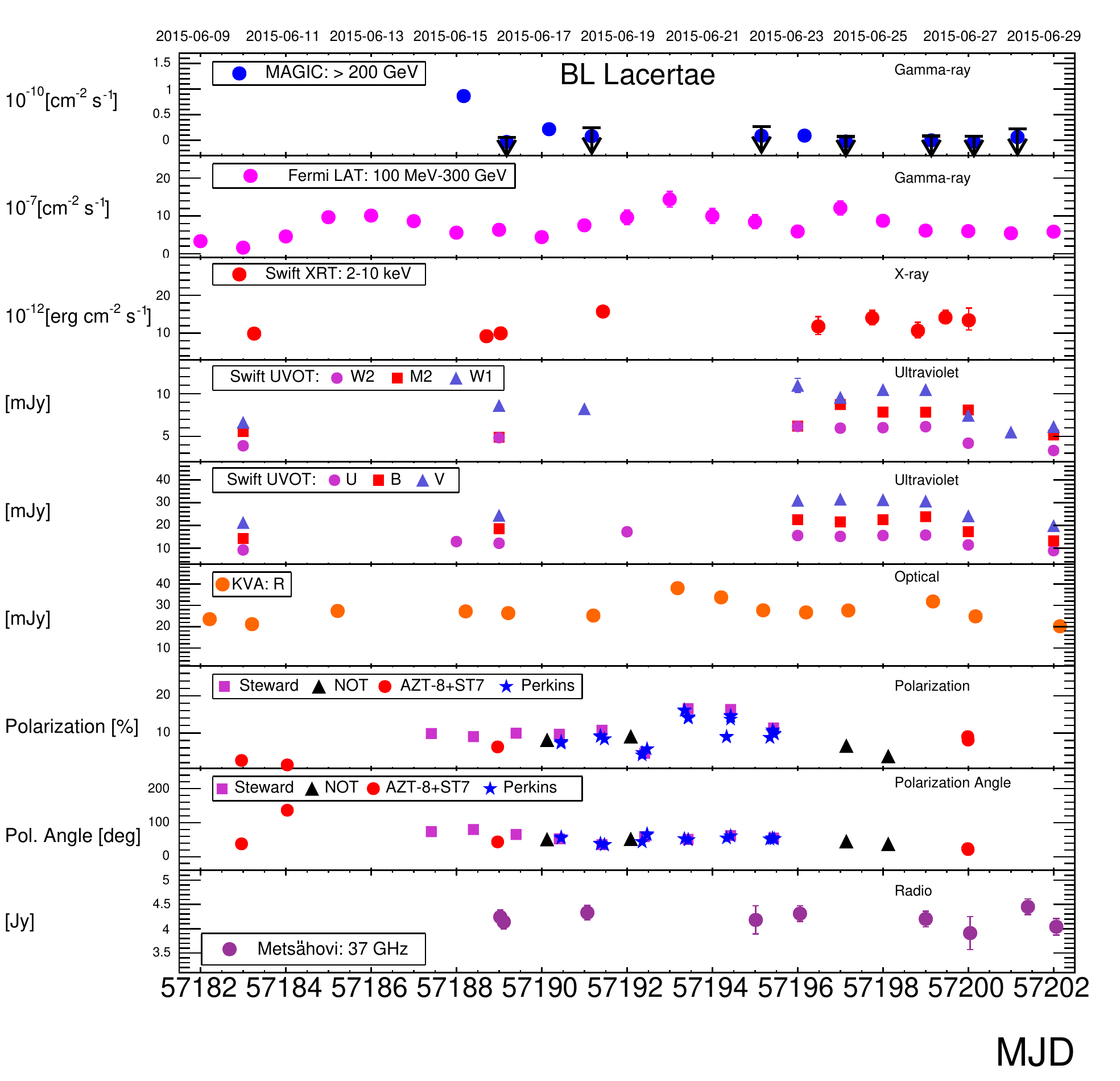}
\caption{Multiwavelength light curves of BL Lac in the range from {\bf 57180 (2015 June 7)} to 57202 (2015 June 29). Data as in Fig.~\ref{MWL_LC_long}.}
\label{MWL_LC_short}
\end{figure*}
     
\section{Results}
\label{section:Results}
\subsection{MAGIC data}
\label{subsection:MAGIC_Results}

Figure \ref{Theta2} shows the squared angular distance ($\theta^2$) distribution between the reconstructed event direction and the source position in the camera and the normalized off-source events. For the complete data set (8.58\,h), we found an excess of 905\,$\gamma$--like events over 2212 $\pm$ 21 background events yielding a significance of 16.4\,$\sigma$ above 70\,GeV within 0.02\,${\rm deg}^2$. The first night alone (2015 June 15, 1.03\,h) showed an excess of 551\,$\gamma$--like events over 234 $\pm$7 background events which yields a significance of 24.8\,$\sigma$ above 70\,GeV within 0.02\,${\rm deg}^2$. The complete dataset without the first night shows a significance of $6.9\,\sigma$. The significances for the individual nights are listed in Table \ref{MAGIC_results}.

Figure \ref{MAGIC_LC} shows the daily light curves of the MAGIC observations, showing the large flare on 2015 June 15. The integral flux above 200\,GeV is $(8.7\pm 0.7)\times10^{-11}$ photons cm$^{-2}$ s$^{-1}$ for the first night and $(4.7\pm 1.5)\times10^{-12}$ photons cm$^{-2}$ s$^{-1}$ for the nights 2015 June 16 to June 28. The average integral flux for the whole period is $(1.5\pm 0.2)\times10^{-11}$ photons cm$^{-2}$ s$^{-1}$. The daily fluxes are reported in Table \ref{MAGIC_results}. In case of non-detection, we calculated 95\% confidence level upper limits of the flux, following \citet{Rolke}, considering a systematic error on flux estimation of 30\% \citep{Aleksic16}.

We also searched for intranight variability on the first night and re-analyzed the data with a binning of $\sim7$ minutes, see inset of Fig.~\ref{MAGIC_LC}. The light curve was fitted with:
\begin{eqnarray}
F(t)=F_0\times2^{-t/\tau}
\end{eqnarray}  
where $F_0$ is the peak flux registered on 2015 June 15 and $\tau$ is the halving time scale, which resulted in $\tau=26\pm8$ minutes.

Figure \ref{MAGIC_SED} shows the spectrum for the whole period (2015 June between 15 and 28). The average spectrum was corrected for the absorption by extragalactic background photons using the model of \citet{do11}. The spectrum can be described by a simple power law ($\chi^2/{\rm d.o.f}=9/10$)
\begin{eqnarray}
\frac{dN}{dE} = f_0\times\left(\frac{E}{300 {\rm GeV}}\right)^{-\alpha}.
\end{eqnarray}
with flux normalization 
$f_0 = (5.6 \pm 0.5)\times 10^{-11}$\,photon cm$^{-2}$ s$^{-1}$\,TeV$^{-1}$
and photon index
$\alpha = 2.9 \pm 0.1$.

The spectrum which was observed on 2015 June 15 is better described by a log parabola even after the EBL correction (LP $\chi^2/{\rm d.o.f}=4.6/9$ vs. PL $\chi^2/{\rm d.o.f}=12.4/10$ )
\begin{eqnarray}
\frac{dN}{dE} = f_0\times\left(\frac{E}{200 {\rm GeV}}\right)^{-\alpha -\beta{\rm log}(E/200 {\rm GeV})}
\end{eqnarray}
with flux normalization 
$f_0 = (3.7 \pm 0.3)\times 10^{-10}$\,photon cm$^{-2}$ s$^{-1}$ TeV$^{-1}$
,
$\alpha = 3.0 \pm 0.2$
and 
$\beta = 0.8 \pm 0.4$.

\subsection{Fermi-LAT}
\label{subsection:Fermi_Results}

\begin{table*}
\begin{center}
  \begin{tabular}{l|c|c|c|c}
    \hline
    \hline
Period            & Date                  & Energy range & Photon index & Flux (10$^{-8}$ ph cm$^{-2}$ s$^{-1}$)  \\
\hline
Total period      & 2015 May 1--July 31   & 0.1--300 GeV & 2.20 $\pm$ 0.03  & 50.1 $\pm$ 1.6 \\
pre-MAGIC period  & 2015 May 1--June 14   & 0.1--300 GeV  & 2.24 $\pm$ 0.04 & 48.0 $\pm$ 2.2 \\
MAGIC period      & 2015 June 15--June 28 & 0.1--300 GeV  & 2.12 $\pm$ 0.04 & 76.8 $\pm$ 5.0 \\
MAGIC period      & 2015 June 15--June 28 & 1.0--300 GeV  & 2.22 $\pm$ 0.11 &  6.0 $\pm$ 0.7 \\
post-MAGIC period & 2015 June 29--July 31 & 0.1--300 GeV  & 2.20 $\pm$ 0.07 & 43.5 $\pm$ 2.6 \\ 
MAGIC detection   & 2015 June 15          & 0.1--300 GeV & 2.29 $\pm$ 0.24  &57.0 $\pm$ 15.4\\
\hline
\end{tabular}
\caption{{\it Fermi}-LAT analysis results for different periods considered (see text). No significant spectral variability was detected. There is no significant change of the photon index during the different sub-periods}
\end{center}
\end{table*}

Integrating over 2015 May 1--July 31  the fit with a power-law model, $dN/dE\propto$ $(E/E_{0})^{-\Gamma_{\gamma}}$, results in TS = 3582 (corresponding to $\sim$60$\sigma$) in the 0.1--300\,GeV energy range, with an integrated average flux of (50.1 $\pm$ 1.6)$\times$10$^{-8}$ ph cm$^{-2}$ s$^{-1}$ and a photon index of $\Gamma_\gamma$ = 2.20 $\pm$ 0.03. In order to test for curvature in the $\gamma$-ray spectrum of BL Lac, an alternative spectral model to the PL, a LP, $dN/dE \propto$ $(E/E_{0})^{-\alpha-\beta \, \log(E/E_0)}$, was used for the fit. We obtain a spectral slope $\alpha$ = 2.13 $\pm$ 0.03 at the reference energy $E_0$ = 347.9 MeV, a curvature parameter around the peak $\beta$ = 0.04 $\pm$ 0.01, and a TS = 3591. We used a likelihood ratio test to check the PL model (null hypothesis) against the LP model (alternative hypothesis). These values may be compared by defining the curvature test statistic: TS$_{\rm curve}$=TS$_{\rm LP}$--TS$_{\rm PL}$=9, meaning that a curved spectral shape is preferred at the 3-$\sigma$ level. 
The $\gamma$-ray light curve of BL Lac for 2015 May 1--July 31 was derived using a log-parabola model and 1-day time bins. For each time bin, the spectral parameters of BL Lac and all sources within 10$^{\circ}$ of it were frozen to the values resulting from the likelihood analysis over the entire period. When TS $<$ 10, 2$\sigma$ upper limits were calculated. The statistical uncertainties in the fluxes are larger than the systematic uncertainty \citep{ackermann12} and only the former are considered in this paper.

The observation period was divided in three sub-periods: pre-MAGIC (2015 May 1--June 14), MAGIC (2015 June 15--28), and post-MAGIC (2015 June 29--July 31) observations. We checked possible spectral changes of the LAT spectrum during the MAGIC observation period with respect to the average spectrum and the spectra collected before and after the MAGIC observation period. In none of the three sub-periods is the LP statistically preferred to the PL model, therefore a PL model is used in the following analysis. We left the photon index free to vary in these three periods and in the night of 2015 June 15 (MJD 57188), at the time of the MAGIC detection of the source at VHE. The results are reported in Table 2.

For the MAGIC period, we investigated the LAT data also with sub-daily time bins. We produced a $\gamma$-ray light curve with 12-hr and 6-hr time bins using a PL model in the 0.1-300 GeV energy range. For each time bin, the spectral parameters of BL Lac and all sources within 10$^{\circ}$ of it were frozen to the values resulting from the likelihood analysis over the entire period, except the normalization was left free to vary. In the following analysis of the sub-daily light curves, we fixed the flux of the diffuse emission components at the value obtained by fitting the data over the respective daily time-bins. No significant flux variability was observed by {\it Fermi}-LAT on sub-daily time bins during the MAGIC period. 
 
Analysing the LAT data collected over 2015 June 15--28 in the 1--300 GeV energy range with a PL, the fit yielded a TS = 496 with a photon index of 2.22$\pm$0.11 and a flux of (6.0$\pm$0.7)$\times$10$^{-8}$ photons cm$^{-2}$ s$^{-1}$. By means of the \texttt{gtsrcprob} tool, we estimated that the highest energy photon emitted by BL Lac (with probability $>$ 90\% of being associated with the source) was observed on 2015 June 21 with an energy of 31.9 GeV. 

\subsection{Multiwavelength light curves}
\label{subsection:MWLLC_Results}

The multiwavelength light curves of BL Lac in 2015 are shown in
Figs.~\ref{MWL_LC_long} and \ref{MWL_LC_short}. Figure~\ref{MWL_LC_long}
  shows an extended period from April to August, while
  Fig.~\ref{MWL_LC_short} shows the zoom {\bf around} the MAGIC observing
  period. The radio flux is rather low during this period in
comparison to its long-term behaviour \citep[see e.g.][]{Nieppola09}
and shows no major outbursts.

 In optical R-band the general flux level is significantly larger
  than the long-term average of 13.1\,mJy \citep{Lindfors16}. During
  the campaign the flux doubles from 20\,mJy to 40\,mJy, reaching the
  largest flux in late 2015 June. The optical U- B- and V-band as
well as the ultraviolet $w2$- $m2$- and $w1$-band light curves follow
the same trend as the R-band, even though observations during the
largest $R$-band fluxes are not available in other bands. In X-rays
the fluxes are also at typical levels for the source. The {\it Fermi}-LAT
light curve shows multiple flares. For June, during which the optical
light curve is well-sampled, the variability seems to be rather
simultaneous in HE $\gamma$-ray and optical bands. The variability in these
two bands is often correlated in this source \citep[see e.g.][]{bloom97,Ramakrishnan16}.

As shown in Sect 3.1 the MAGIC light curve shows very high flux on 2015
June 15. We have simultaneous or quasi-simultaneous data from {\it Fermi}-LAT,  {\it Swift}-XRT and KVA, but none of the bands show an increased
flux. In particular the {\it Fermi}-LAT 6\,h and 12\,h light
curves do not show any significant variability during or around this
period. The X-ray data from {\it Swift}-XRT are not strictly simultaneous, but were taken $\sim$0.5\,days later, and taking into account the fast
variability in the VHE band, we cannot exclude that a fast flare also
happened in X-rays.

\subsection{Optical polarization behaviour}

 We combined the optical polarization observations from five telescopes to investigate the behaviour of optical polarization degree and electric vector position angle (EVPA) during the extended period of activity. As the EVPA has $\pm180^o \times n$ (where $n=$1, 2,...) ambiguity, we selected the values such that the differences between any two positions are minimised. There was one data point (MJD 57184.5) which differed by $\sim 90^o$ from the previous observation and can therefore be either $-43.5^o$ or $+136.5^o$, we have plotted it as $+136.5^o$. 

  BL Lac is known to show highly variable degree of polarization ($<1$\% to 40\% \citet{Hagen-Thorn02}) in timescales of tens of minutes \citep{Covino15}. During our observations the optical polarization degree varies between 1-20\%, which is within the typical range for this source. Four nights before the detection of the VHE $\gamma$-ray flare the polarization degree was very low{\bf,} $1.4\%$, while on the night of the flare it was $9\%$.

There are two rotations of the EVPA. First rotation starts around MJD 57161 and ends around MJD 57175, few days before the start of the VHE $\gamma$-ray observations, but simultaneous with increasing flux in the {\it Fermi}-LAT energy range. The rotation is from $-13.8^o$ to $95.1^o$. Length and starting time of the second rotation depends on the data point of MJD 57184.5. Independent of whether we correct the data point of MJD 57184.5 or not (see above), a rotation of EVPA is observed during the VHE $\gamma$-ray flare. In the first case EVPA rotates from $80^o$ to $-27^o$ in 27 days with the rotation
starting on the night of the detection of the VHE flare. In the second case EVPA rotates from $137^o$ to $-27^o$ in 31 days and the rotation starts 5 days before the detection of the VHE flare.
Neither rotations are very smooth and they are in opposite directions. Rotations with $>90^o$ are rather common in the source, e.g. \citet{Jermak16} identified four such rotations from their dataset covering four years of data (2008-2012). For single events it is therefore difficult to conclude on the connection of the rotation with the VHE $\gamma$-ray flare.
 
\begin{figure}
  \centering
  \includegraphics[width=0.43\textwidth]{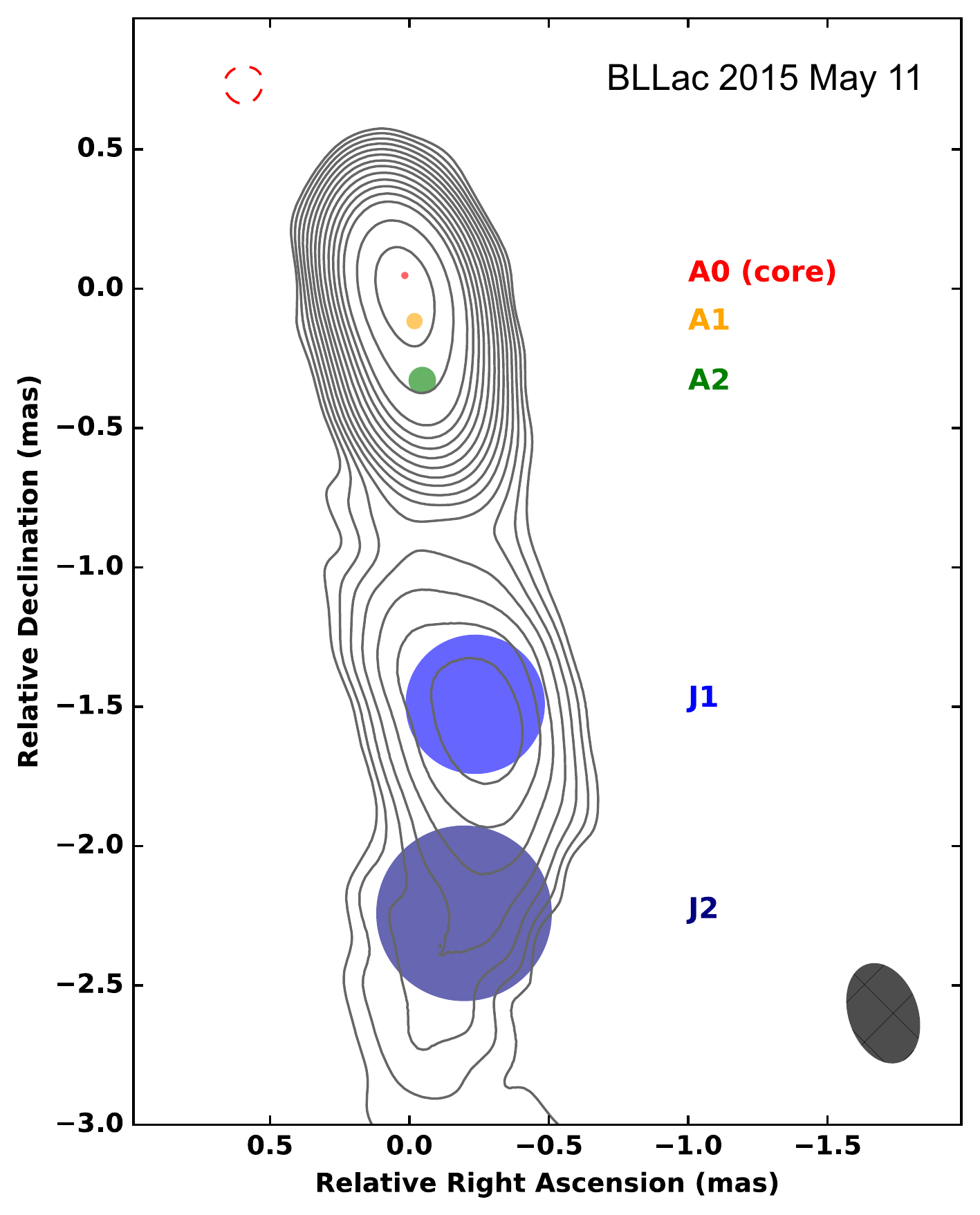}
  \caption{A 43\,GHz VLBA image of BL Lac on 2015 May 11. Contours start from -1, 1, 2,...times 1.7mJy/beam and increase by factors of $\sqrt{2}$ (negative contours are colored in red). The beam size is shown at the bottom-right corner of the plot and is 0.325 mas x 0.208 mas at PA 20.6 degrees. The peak flux density of the map is 1.88 Jy/beam. The image also superposes a set of best-fitted circular Gaussian components - $A0$(=core), $A1$, $A2$, $J1$ and $J2$. $A0$, $A1$ and $A2$ are well-known stationary features located at the core, $\sim$0.14\,mas and $\sim$0.3\,mas away respectively\citep[see e.g.][]{Wehrle16}}. 
  \label{vlba_jet}
\end{figure}

\begin{figure}
  \centering
  \includegraphics[width=0.95\textwidth, angle=-90]{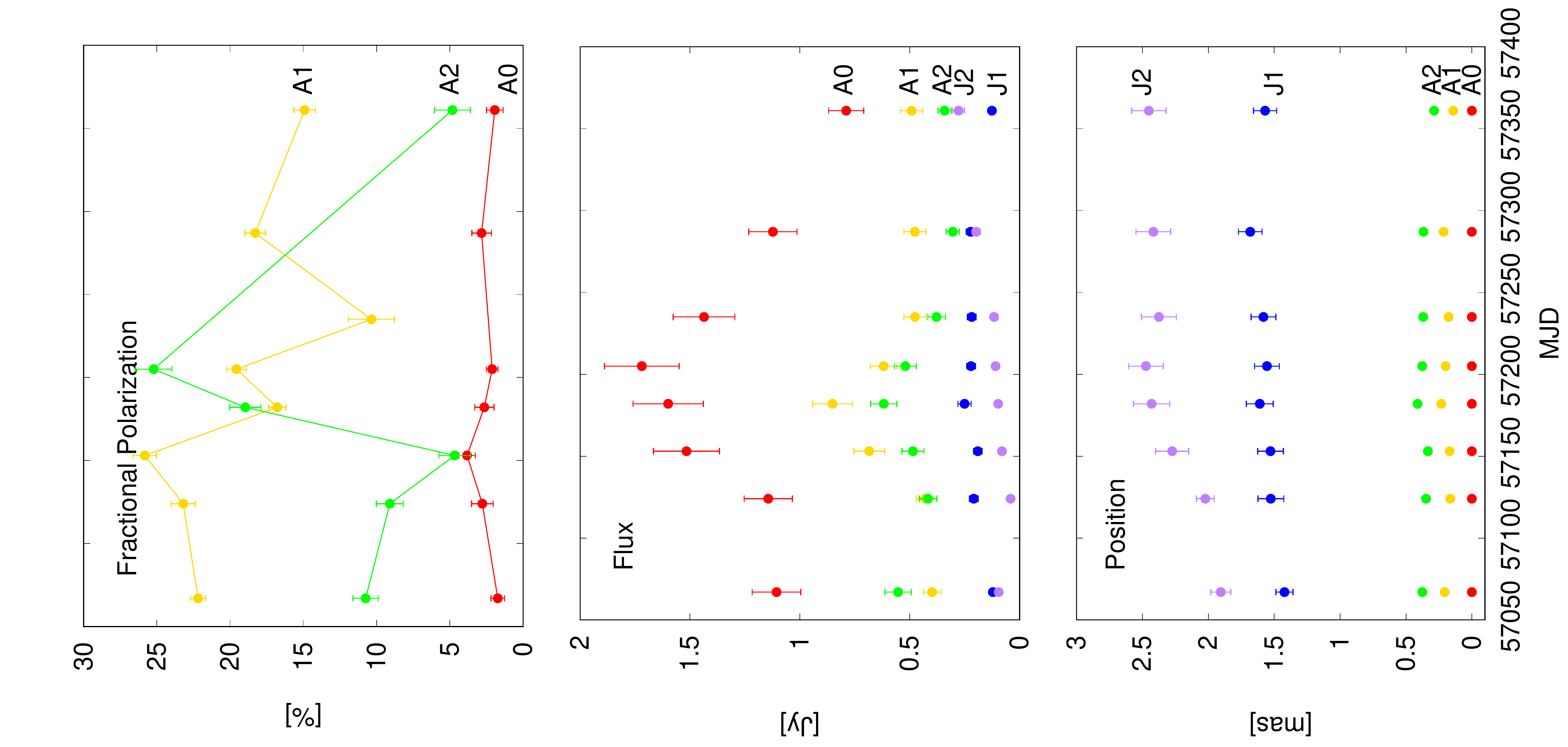}
  \caption{Evolution of polarization fraction (top), flux density (middle) and position (bottom) of the VLBA components as function of time. We adopt a flux measurement uncertainty of 10\%, which is typical for VLBA data, and the position error is estimated to be one-fifth of the fitted Gaussian size. As the polarized intensity images in MJD 57235 and 57287 have a higher noise level (see Section 2.5), polarization fraction could not be derived for all components in these epochs.}
  \label{vlba_pos}
\end{figure}

\begin{figure*}
  \centering
  \includegraphics[width=0.85\textwidth]{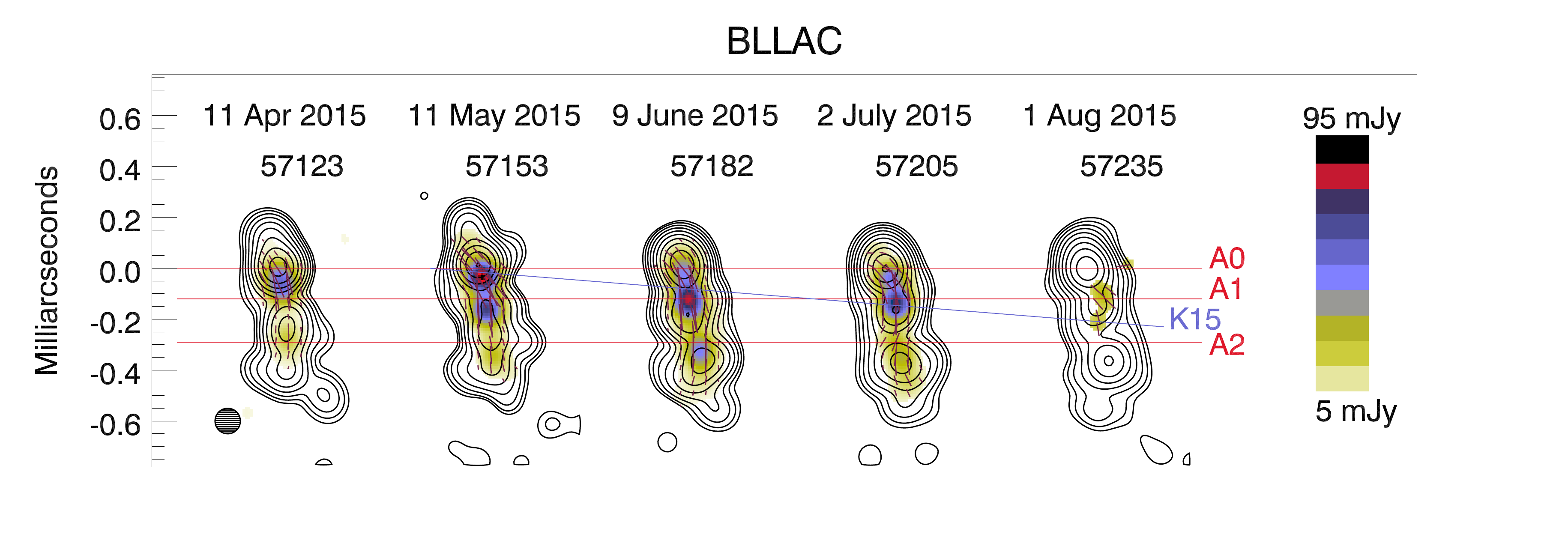}
  \caption{A series of five 43 GHz total (contours)  and polarized (color scale) intensity images of BLLac with a resolution beam of 0.1x0.1mas and the total intensity
peak of 1.59 Jy/beam 
The contours are 0.4, 0.8, 1.6, 3.2, 6.4, 12.8, 25.6, 51.2\% of the peak plus one additional contour 96\% of the same peak to localize the peak more precisely). Red linear segments within images indicate direction of polarization;
the red horizontal line mark positions of the core A0, and stationary features, A1 and A2, according to \citet{jorstad17}. The blue line
indicates a possible motion of a hypothetical knot K15.}
  \label{vlba_pol2}
\end{figure*}

\subsection{VLBA 43\,GHz results}
\label{subsection:VLBA_Results}

We investigated the jet kinematics of BL Lac in 2015 using eight epochs
of 43\,GHz VLBA images. The image of 2015 May 11 (MJD 57153) is shown in
Fig.~\ref{vlba_jet}. It shows the position of the components $A0$, $A1$, $A2$, $J1$ and $J2$.

We do not find any clear emergence of new components over the examined
period. For all the epochs, the images were similar to that presented in Fig.~\ref{vlba_jet}. The positions of the components at different epochs are shown in Fig.~\ref{vlba_pos} (bottom panel), which reveal no successive motion of $A1$ and $A2$ with respect to $A0$.
These three components are well-known stationary
components \citep[see e.g.][]{Wehrle16} located at the core,
  $\sim$0.14\,mas and $\sim$0.3\,mas from the core. For BL Lac 1\,mas
  corresponds to 1.3\,pc in projected distance. Adopting a jet viewing
  angle of $6^o$ \citep{Wehrle16}, the deprojected scale results in 1\,mas$\sim$13\,pc. So, the components $A1$ and $A2$ are at the distance of
  1.8\,pc and 3.9\,pc. Knots $A1$ and $A2$ are imaged as a single feature at 15\,GHz and have been interpreted as a quasi-stationary recollimation shock \citep{Cohen14}. As discussed in \citet{jorstad17} the stationary features make a detection of moving knots difficult.

While there were no clear ejections, a flux density increase of the
core or of the central one-beamsize region is measured from 2015 May to July, as can be seen in Fig.~\ref{vlba_pos} (middle panel). This brightening could be due to a new moving feature (knot).
There is also an increase of polarized flux density in the core in May (57153), then in A2 in June and July (57182, 57205), visible in Fig.~\ref{vlba_pos} (top panel). Also Fig.~\ref{vlba_pol2} presents variability of the degree of polarization of the core and stationary features. It shows an increase of the fractional polarization in the core and A1 in May (57153), followed by an increase of the degree of polarization in A2 in June (57182). An increase of fractional polarization along with the total flux density is usually interpreted as the result of a shock propagation in the jet \citep{hughes}. Therefore, also polarization observations show a hint of a new moving feature.

If we interpret this as a moving knot, which could cause such a behaviour, it moves with a proper motion
of $\sim$1.07mas/yr ($\sim$5c), fairly common for BL Lac (see Fig.~\ref{vlba_pol2}). 

This brightening and increase of polarized flux intensity and degree
of polarization are in line with the general brightening of the source
in optical and $\gamma$-ray bands and therefore the knot is a likely
location of the activity. As it is relevant for the SED, we also
estimate the distance from the central black hole for this emission
region.  The brightening is within 0.2\,mas from the radio core (or
might be the core itself), which corresponds to 2.6\,parsecs.
Additionally we have to take into account the location of the radio
core relative to the black hole. For BL Lac, \citet{Pushkarev12}
suggest that the deprojected separation between the black hole and
the 15\,GHz core to be 0.84\,pc based on their core-shift
measurements. If we assume that the radio core position is inversely
proportional to frequency \citep{OSullivan09,Pushkarev12}, the
separation between the black hole and the 43\,GHz core would be
$\sim$0.3\,pc. This gives us a range of distances from $\sim$0.3\,pc
to $\sim$2.9\,pc from the central black hole.

\section{Comparison with other VHE flares from BL Lac}

In addition to the flare presented here, VERITAS has observed bright,
fast VHE $\gamma$-ray flares from this source on two occasions
\citep{Arlen13,2018arXiv180210113A}. In the following we compare the VHE $\gamma$-ray and multiwavelength behaviour during these three epochs:
2011 June (hereafter VERITAS flare 1), 2015 June (hereafter MAGIC
flare) and 2016 October (hereafter VERITAS flare 2). Observations of the VERITAS flare 1, similarly to the MAGIC flare, were triggered by high state in $\gamma$-ray and optical bands. On the contrary, VERITAS flare 2 was observed as part of the regular monitoring program.

The maximum VHE $\gamma$-ray flux we observed in the night of 2015 June
15 is $(1.5\pm 0.3)\times 10^{-10}$ photons cm$^{-2}$ s$^{-1}$. This is
only half the flux VERITAS observed from the source on
2011 June 28 (VERITAS flare 1), but was the brightest VHE flux from the source since that observation. However, on 2016 October 5 VERITAS detected yet another
bright VHE $\gamma$-ray flare from the source (VERITAS flare 2) \citep{icrc_veritas,2018arXiv180210113A} with a peak flux
of $(4.2\pm0.6)\times10^{-10}$ photons cm$^{-2}$ s$^{-1}$, which is
also significantly brighter than the flux we report. Like in VERITAS
flare 1 and 2, also for the MAGIC flare the signal is
concentrated on one night. The observations from the following nights do
not show significant signal.

During the MAGIC flare the flux decayed with halving time scale of
$26\pm8$ minutes. In VERITAS flare 1, the halving time (also
corresponding to decay of the flare) was approximately a factor of two
shorter: $13\pm4$ minutes. For the VERITAS flare 2, the rise time was
$140^{+25}_{-11}$ minutes and the decay time $36^{+8}_{-7}$
minutes. The value is compatible within the uncertainties with the decay time we
measure for the MAGIC flare.

All three detected occasions of fast VHE $\gamma$-ray variability
(MAGIC flare and VERITAS flares 1 and 2) have occured during an
extended period (lasting some weeks) of high flux in the HE
$\gamma$-ray band ($F_{E>0.1-300\,GeV} >0.5\times10^{-6}$ photons
cm$^{-2}$ s$^{-1}$), even if the $F_{E>100\,MeV}$ $\gamma$-ray flux in
the nights of the high VHE $\gamma$-ray flux is not particularly
high. The same is true for the optical flux; while long term average
R-band flux from Tuorla blazar monitoring program is 13.1\,mJy
\citep{Lindfors16}, the optical flux during these three epochs has
been between $20-40$\,mJy. On the contrary, the X-ray flux did not
show any long term high states during any of the three flaring
epochs. Note that high optical and $\gamma$-ray states can be due to an observational bias, because the high states in these bands were used to trigger the observations in the VHE $\gamma$ ray band by MAGIC and by VERITAS (flare 1).

In all three cases the polarization degree drops to rather low values during or just before the observation of the VHE $\gamma$-ray flare.
As discussed in Section~\ref{subsection:MWLLC_Results}, rotations of the optical
polarization angle are rather common in this source and seem to have
occured around the time of the three VHE $\gamma$-ray flares. The rotation during
VERITAS flares 1 and 2 and the MAGIC flare are not particularly smooth
nor long, unlike the rotation shown in \citet{marscher08}. 
Therefore it is very difficult to conclude on a connection
between the optical polarization behaviour and the VHE $\gamma$-ray flares
in BL Lac and more data are certainly needed.
However, the observed behaviour (drop in polarization degree and rotating EVPA) is in agreement with the model of \citet{marscher14} as suggested by \citet{icrc_veritas,2018arXiv180210113A}. In this model, the VLBA core is interpreted as a conical shock through which turbulent shells of plasma pass creating an environment for an efficient acceleration of electrons.

The radio fluxes in 15\,GHz or 37\,GHz are not particularly high
during the onset of the $\gamma$-ray flaring activity. However, around
all the three epochs some activity was detected in 43\,GHz VLBA data;
during the VERITAS flares 1 and 2, a new component was ejected from
the core.
For VERITAS flare 2 the ejection of the component is
  tentative \citep{icrc_veritas,2018arXiv180210113A} and the case observed for the MAGIC flare is very similar to that one. There are clear indications of new component, but due to the standing features very close to core, the analysis is inconclusive. Overall, our observations are in agreement with the connection between the VHE $\gamma$-ray flares and the activity in the 43\,GHz VLBA core
suggested in \citet{icrc_veritas,2018arXiv180210113A}.

It is very suggestive that all three VHE $\gamma$-ray flares show similar multiwavelength variability patterns. However, as discussed above, there are significant uncertainties and observational biases that must be understood with more observations before any firm conclusions on such patterns can be made.

\begin{figure}
  \centering
  \includegraphics[width=0.44\textwidth]{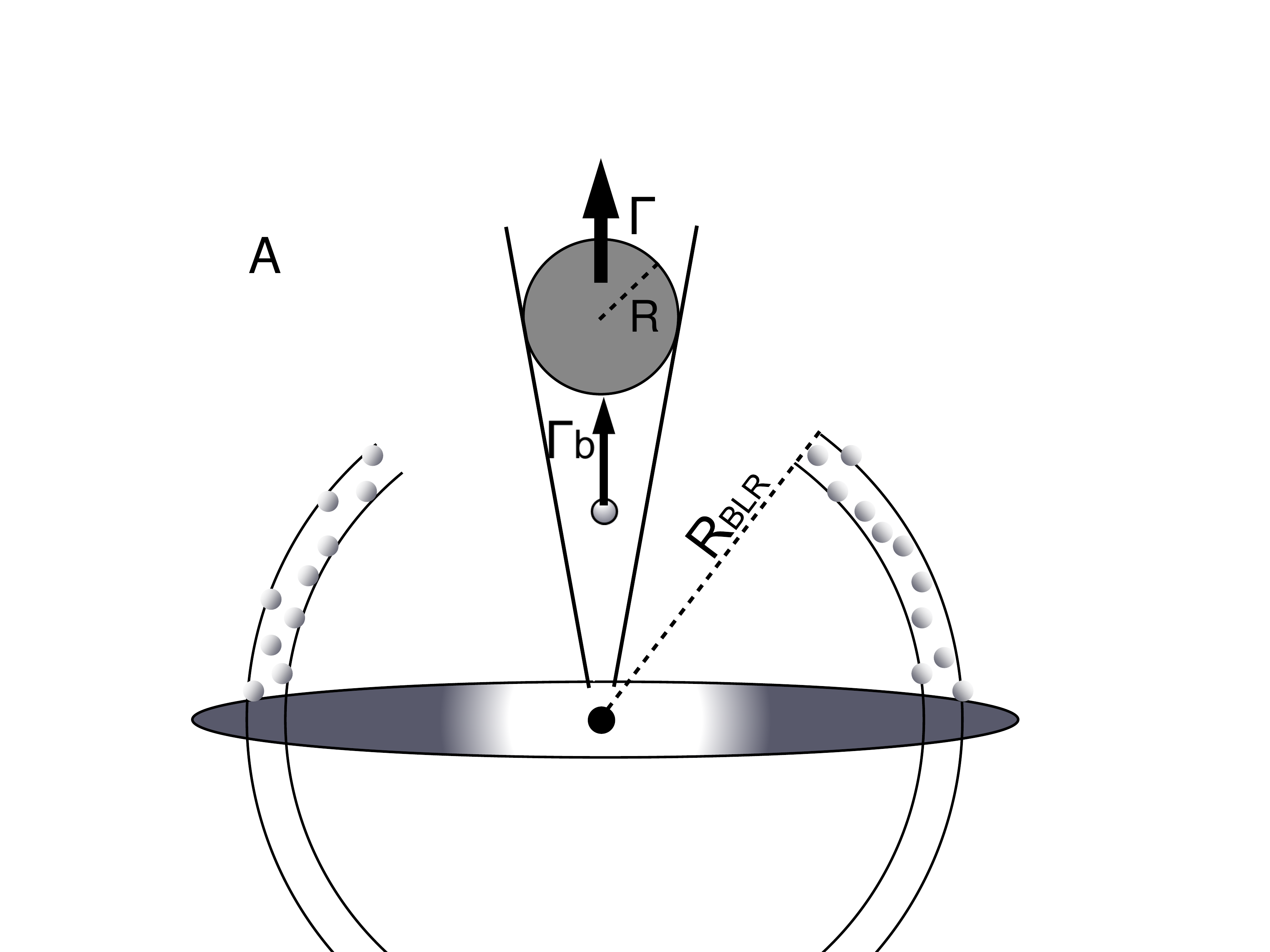}
  \includegraphics[width=0.44\textwidth]{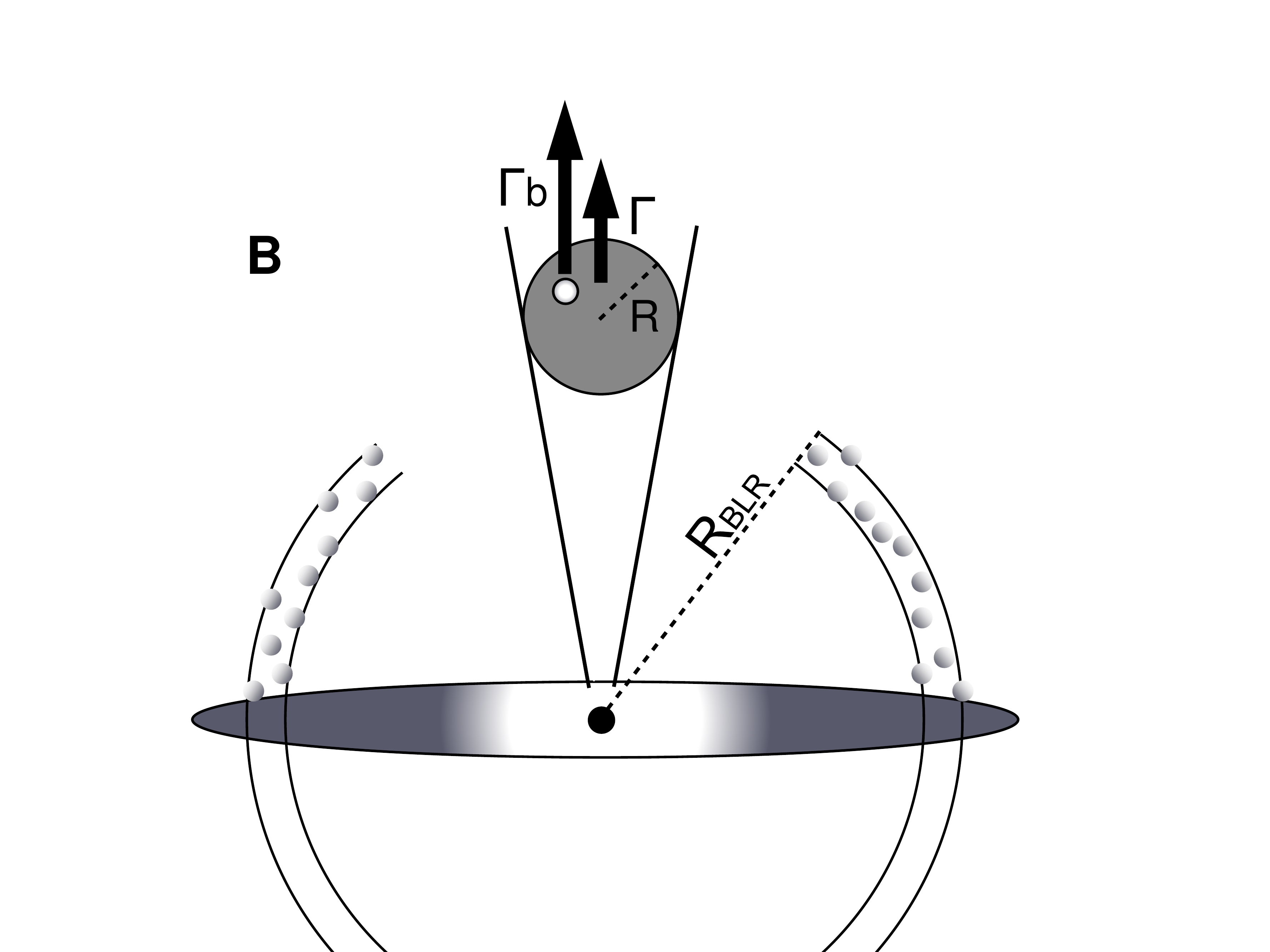}
     \caption{Sketch of the two geometrical setups adopted in the SED modellings. SED is modelled with two emission zones where the smaller region (blob) is located inside the BLR (A) or interacting with the larger region (B), the radio core.}
  \label{cartoon}
\end{figure}

\begin{figure*}
  \centering
  \includegraphics[width=0.43\textwidth]{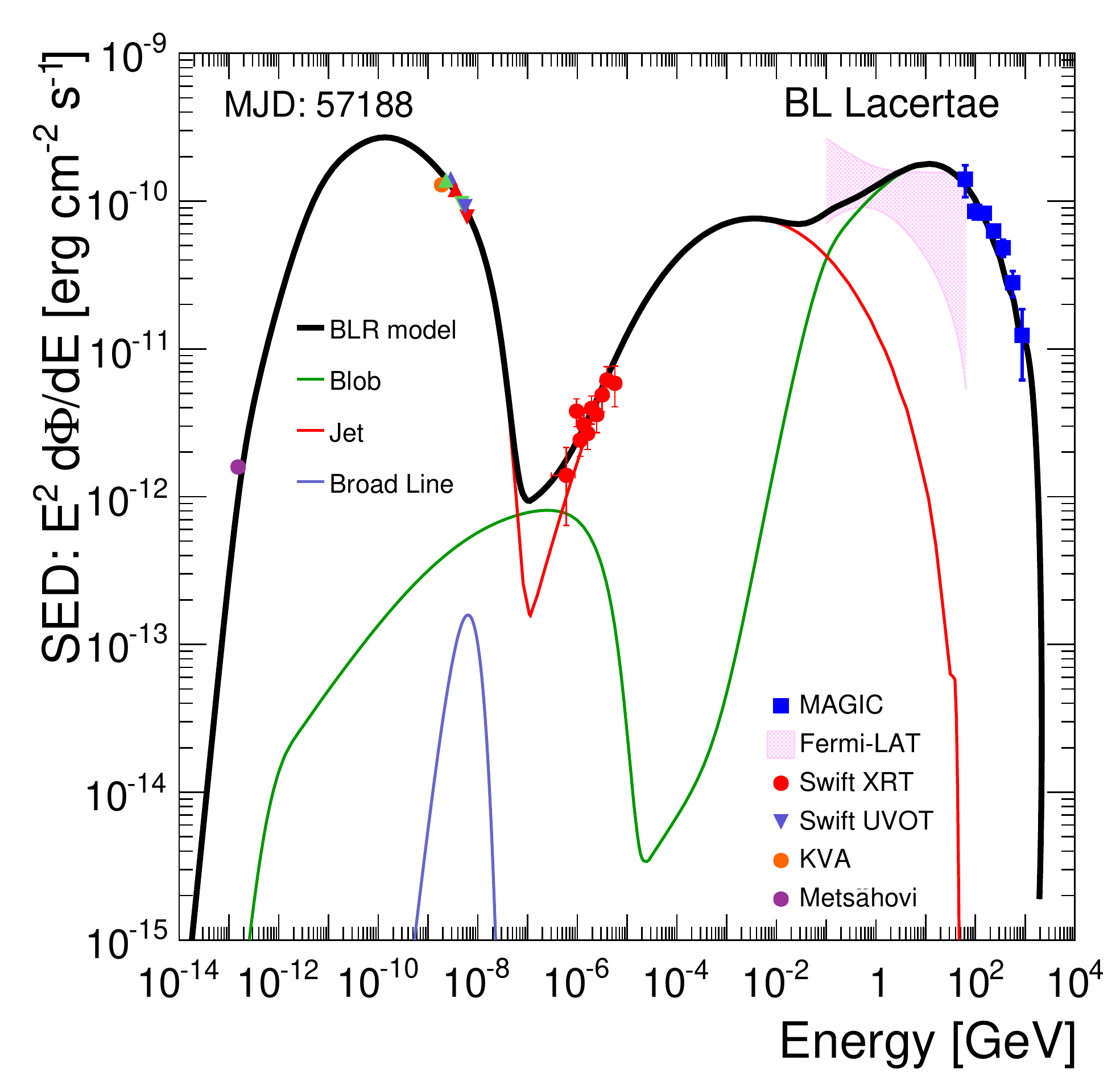}
  \includegraphics[width=0.43\textwidth]{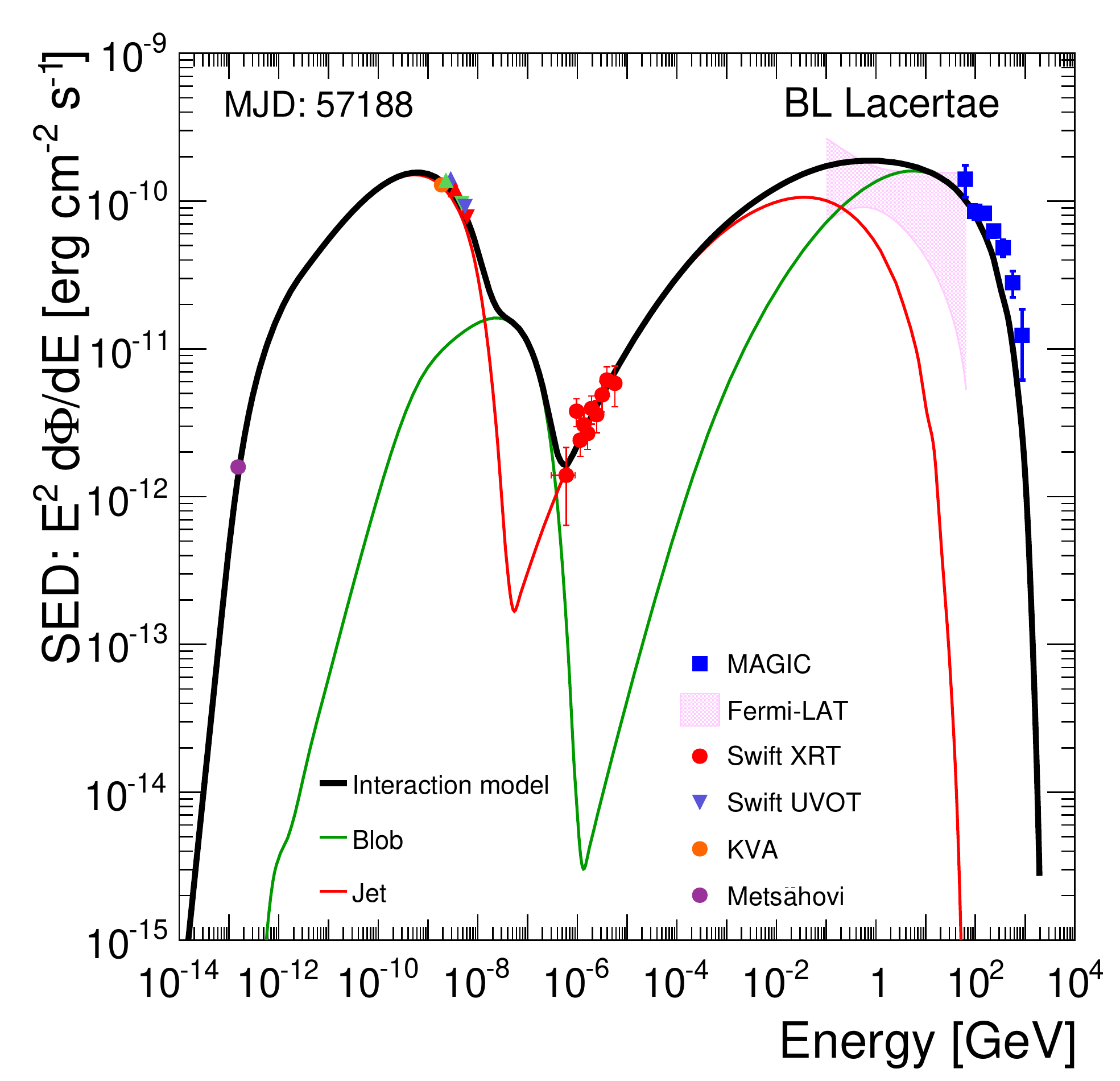}
  \caption{Multiwavelength SED of BL Lac on MJD 57188. 
    SED is modelled with two emission zones where the smaller region (blob) is located inside BLR (left) or interacting with the larger region (right).}
  \label{MWL_SED}
\end{figure*}

\section{Multiwavelength SED modelling}
\label{section:MWLSED_Results}
The SED of BL Lac has been modelled
several times in the past. First ones were homogeneous,
one-zone synchrotron self Compton models \citep[see e.g.][]{ghisellini99,Ravasio02}. However, it was noted already in
the EGRET era that modelling the high flux states above 100 MeV
required external seed photons for Compton scattering
\citep{sambruna99,madejski99,bottcher00}. Ever since, the SED of BL Lac has been conventionally modelled with an
external Compton model \citep[see e.g.][]{bottcher13} using external
photons from the BLR as seed photons.

\citet{bottcher13} also considered a hadronic model to
describe the SED. As is generally the case,
also for BL Lac the hadronic model would require magnetic field
strength of $\sim10$\,Gauss and very large power in relativistic
protons.

The early SED modellings did not include
the VHE $\gamma$-ray data, but it was shown in \citet{Albert07} that
the one-zone SSC model of \citet{Ravasio02} could also describe the
observed VHE $\gamma$-ray data. However, the observed VHE $\gamma$-ray
flux in 2015 is much larger ($\sim$ 10 times) than the one presented in \citet{Albert07}
and in addition a fast variability gives additional constraints to the
model. Very recently, \citet{morris18} presented first SED modelling attempt, including the VHE $\gamma$-ray data from fast flare of 2016 (VERITAS flare 2). The model computes  the  time  evolution  of  a
  reconnecting plasmoid whose radius and velocity evolve as it travels through the reconnection layer. The model can produce the profile of the fast flare, but overproduces the optical to X-ray part of the SED very significantly.

In the following we consider different models to account
for the observed SED and the variability
patterns discussed in Sect.~\ref{subsection:MWLLC_Results} and Sect.~\ref{subsection:VLBA_Results}.

\begin{table*}[th]
\centering
\begin{tabular}{lcccccccccccc}
\hline
\hline
model&component&$\gamma _{\rm min}$ & $\gamma _{\rm b}$ & $\gamma _{\rm max}$ &
$n_1$ & $n_2$ &$B$ & $K$ &$R$ & $\delta$ \\
& &$10^2$&$10^4$&$10^5$ &  & &[G] & [$10^3$cm$^{-3}]$  & $[10^{16}$cm]&\\ 
\hline
BLR&Blob&       1.0&	1.0&	2.0&	 2.0&	3.0&	 0.14&	 45&	0.1&	25\\
&Jet&	5.0&	0.3&	0.3&	1.9&	3.9&	0.12&   0.4	&	30&	7\\
\hline
Interaction&Blob&       50.0&	4.0&	0.9&	2.0&	3.2&	0.013&	300&	0.17&	60\\
&Jet&	3.0&	0.9&	0.3&	2.0&	3.7&	0.05&   0.8	&	30&
7\\
\hline
\hline
\end{tabular}
\vskip 0.4 true cm
\caption{Model parameters for the two SED models of the flare night (MJD 57188): BLR (external photons for IC scattering provided by the BLR) and Interaction (Jet component providing the seed photons for the IC scattering). 
The following quantities are reported: the minimum, break, and maximum
Lorentz factors and the low and high energy slope of the electron
energy distribution, the magnetic field intensity, the electron density, the radius of the
emitting region and the Doppler factor.}
\label{model}
\end{table*}

\subsection{Two-zone modelling of the SED}
We reconstructed the multi-band SED of BL Lac for the flare night (MJD 57188, 2015 June 15) only ($\pm 0.5$ days).

The observed common trends in HE $\gamma$-rays, optical and 43\,GHz radio
core suggest that the emission in these bands comes from the 43\,GHz
radio core. On the other hand, the VHE $\gamma$-ray emission shows fast
variability and must originate from a very small component for which we
do not have constraints on the location from the light curves. For these two components, we adopt a leptonic model similar to one presented in
\citet{tavecchiobecerra}, which assumes two emission components: a
small blob, emitting the rapidly variable VHE emission, and a larger jet (which in our case is the 43\,GHz VLBA core, see above)
responsible for the slower variability in the other bands (see Fig.~\ref{cartoon}). In the model both of these regions are filled with electrons distributed in energy according to a smoothed broken power law:
\begin{equation}
N(\gamma)=K\gamma^{-n_1}\left(1+\frac{\gamma}{\gamma_{\mathrm b}}\right)^{n_1-n_2},   \gamma_{\mathrm min}<\gamma<\gamma_{\mathrm max}.
\end{equation}
The distribution has normalization $K$ between $\gamma_{\mathrm min}$ and $\gamma_{\mathrm max}$ and slopes
$n_1$ and $n_2$ below and above the break, $\gamma_{\mathrm b}$ \citep{2003ApJ...593..667M}. The emission regions have magnetic fields $B$, sizes $R$
and Doppler factors $\delta$ for which we looked for constraints from
observations:

\begin{itemize}
\item The MAGIC observations on 2015 June 15 show variability timescale of 26 minutes, which constraints the size of the emission region to be $R \sim ct_{var}\delta\sim 10^{15}$\,cm for the blob. 
\item  For the larger region, we estimate the size of the emission region from the variability timescale of optical and $\gamma$ light curves (time scale is order of 2 days) to be $10^{17}$\,cm.
\item Magnetic field strength and Doppler factor can be constrained from the VLBA observations. For the magnetic field we use 0.11\,G \citep{Pushkarev12}. For the Doppler factor we use $\delta$=7 \citep{jorstad05,Wehrle16,jorstad17}. We use these values for the larger emission region, i.e the VLBA core.
\end{itemize}

\citet{tavecchiobecerra} suggested two different geometrical
arrangements for the positions of the small and large emission regions:
i) co-spatiality outside the BLR, and ii) a geometry where the larger
emission region is inside the BLR and the small emission
region is outside the BLR to avoid the $\gamma-\gamma$ absorption of VHE $\gamma$ rays. In BL Lac, however, the observed
emission lines are rather weak (L$_{H_\alpha}=4\times10^{41}$ erg/s)
\citep{corbett96,corbett00,capetti10}, which gives us, using the
standard scalings L$_{\mathrm{BLR}}=2.5\times10^{42}$ erg/s and
R$_{\mathrm{BLR}}=2\times10^{16}$cm $\sim 0.005$pc \citep[see e.g.][]{ghisellini09}.
The scaling relation is based on very luminous quasars and therefore not directly applicable to fainter objects like BL Lac, but the estimated size
is at least orders of magnitude smaller than the distance of the 43\,GHz VLBA core to the black hole (at least 0.3\,pc see Sect.~\ref{subsection:VLBA_Results}), therefore, the large emission region is clearly outside the BLR see Fig.~\ref{cartoon}). Another source of external photons that is nowadays commonly
considered in the case of FSRQs is the dusty torus \citep{2008ApJ...675...71S},
but there is no observational evidence of the existence of such
structure in the lower luminosity objects such as BL Lac. Therefore no
external seed photons are considered for the large emission region.

For the small region we do not have constraints on the location from
the variability patterns. In general, as discussed in
\citet[e.g.][]{bottcherandels16,2017MNRAS.464..152A}, in order to
avoid significant $\gamma$-$\gamma$ absorption the VHE $\gamma$-ray emission
region must be located near the outer boundary of the BLR or beyond
it. The BLR in BL Lac is too weak (see above) to absorb the
VHE $\gamma$-rays, but can provide additional seed photons
to the Compton scattering.
Therefore, in the first setup, we consider the small blob to be located
inside the radius of the BLR (see Fig.~\ref{cartoon} panel A). Our modelling represents only one possible set of parameters; but all
observational constraints have been taken
into account and the
 model well reproduces the observed SED (see Fig.~\ref{MWL_SED} left panel). However, if the BLR is indeed as small as suggested by the scaling relation (see above), then in 30 minutes the blob would travel a distance greater than the size of the BLR. Again, as the size of the BLR is very uncertain, we still consider the model feasible.

The other possible model setup is that the two emission regions are
co-spatial and interact with each other, the larger (jet) providing
additional seed photons for inverse Compton scattering (see
Fig.~\ref{cartoon} panel B). To model this case, we adopt the case B
setup of \citet{tavecchiobecerra}, but without external seed photons
from the torus. The same model was used in \citet{0716}. In order to
reproduce the high flux in VHE $\gamma$-rays, we have to use {\bf $\sim50\%$} lower magnetic field than suggested by \citet{Pushkarev12}
based on the VLBA measurements and assuming equipartition between the
energy carried by particles and the magnetic field. The set of
parameters shown in Table \ref{model} reproduces the SED acceptably
(see Fig.~\ref{MWL_SED} right panel), but:
\begin{itemize}
\item  As the shape of the highest energy component derived with the interaction model is wide (similar to SSC) it is difficult to fit the highest MAGIC spectral points without overproducing the {\it Fermi}-LAT spectrum.   
\item It would require
deviation from equipartition condition at least for the smaller component ($U_B\sim 0.04\times U_e$) as is generally found for BL
Lacs when one-zone models are considered \citep{2016MNRAS.456.2374T}.
However, unlike the conclusions in that paper, the deviation from
equipartition is required even in the two-zone modelling taking into
account the interaction between the two zones. We suggest that this is
due to an extreme flaring activity that took place in BL Lac during
the epoch considered here.
\end{itemize}
Again the modelling uses a single parameter set only out of many different possibilities.

\begin{figure}
\centering
\includegraphics[width=0.43\textwidth]{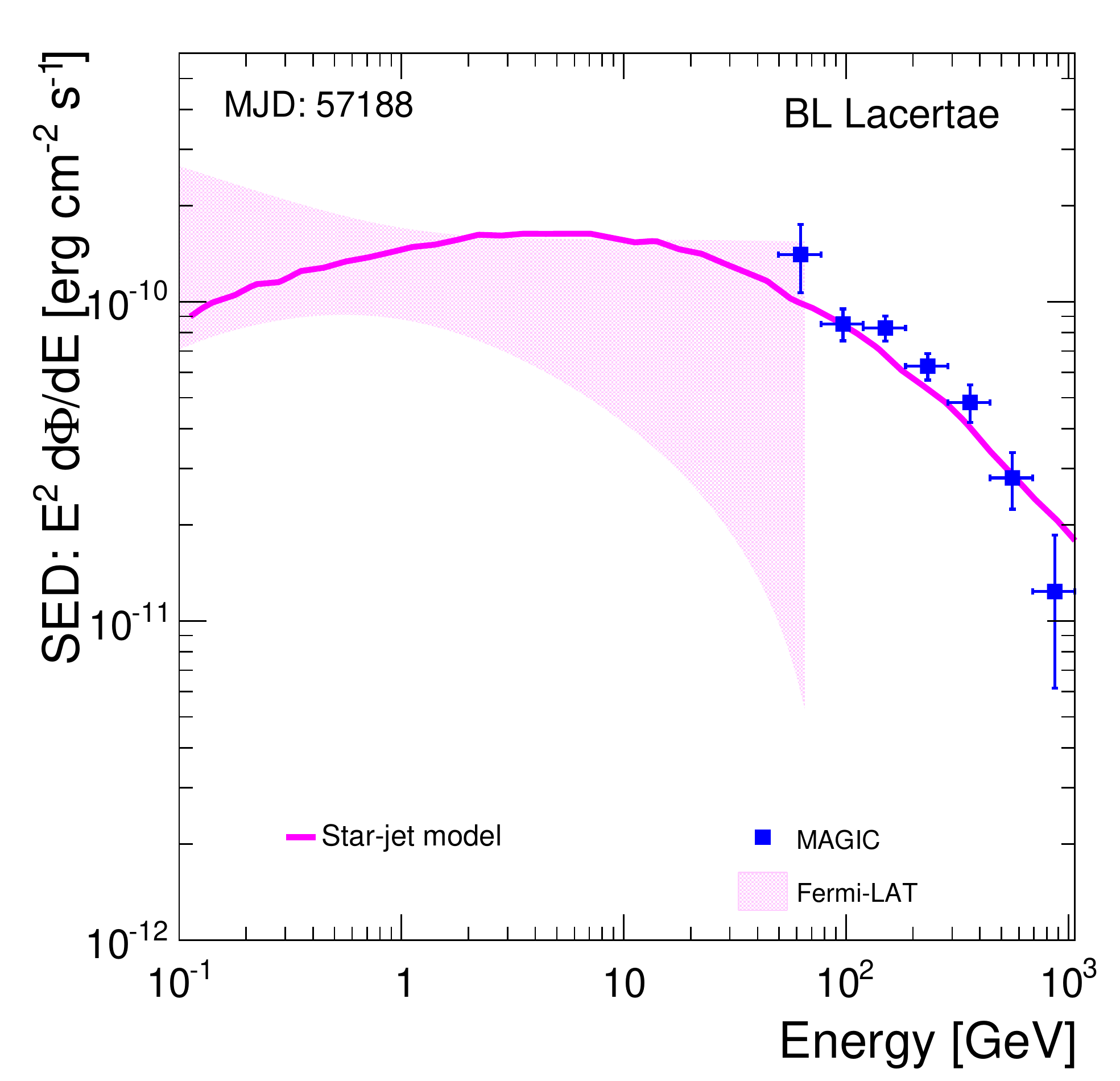}
\includegraphics[width=0.43\textwidth]{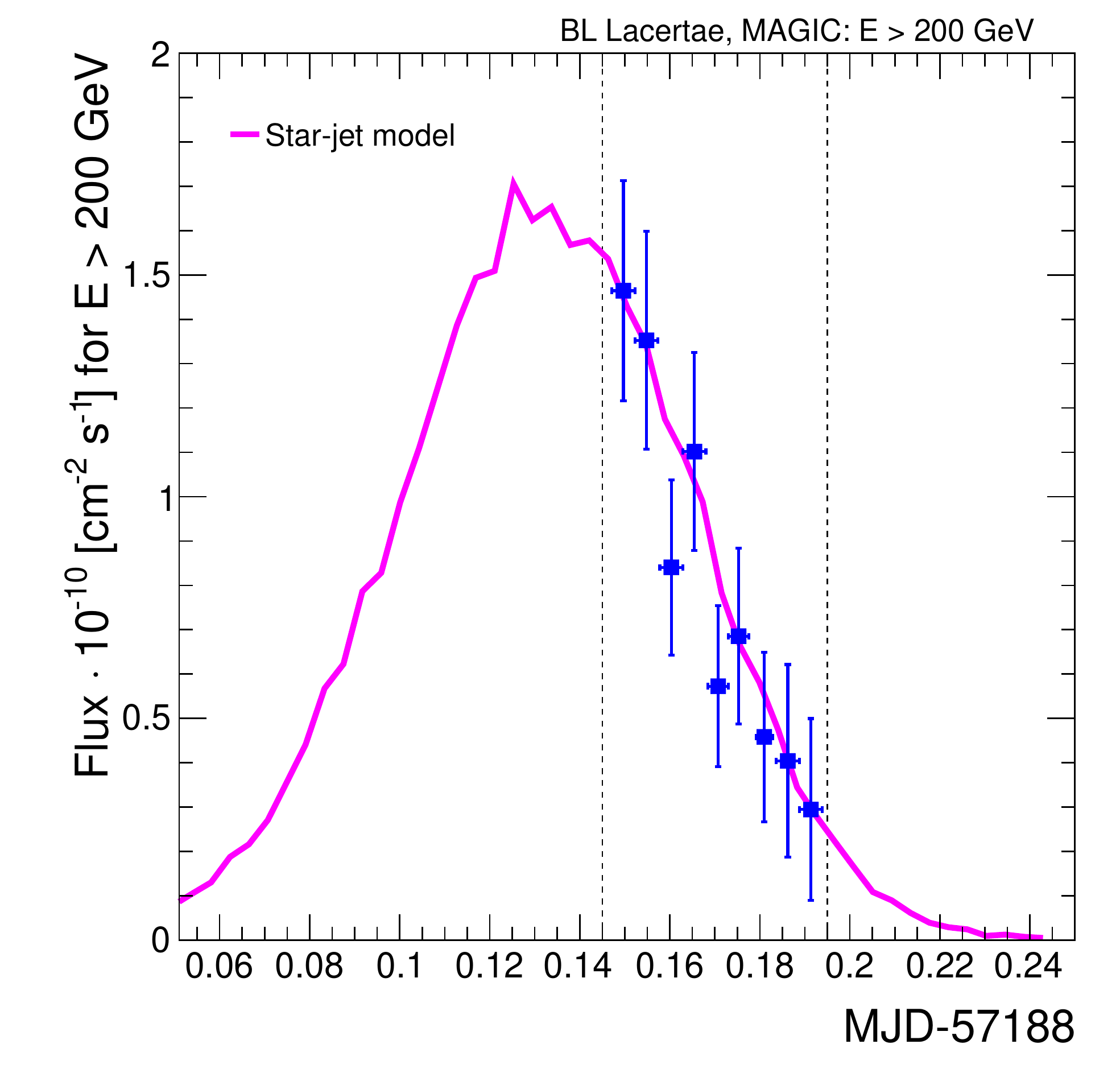}
\caption{
{\it Top panel:} Interpretation of the $\gamma$-ray emission observed by \textit{Fermi}-LAT (shaded region) and MAGIC (full circles) in the framework of the interaction of the relativistic electrons in a blob with the radiation field of a star that has entered the jet (magenta model line). 
The absorption in the Extragalactic Background Light is taken into account according to \citet{do11} model.
{\it Bottom panel:} The MAGIC light curve above 200 GeV.
The dashed vertical lines are the time range from which the spectral energy distribution (top panel) is computed.
}
\label{fig:star}
\end{figure}

\subsection{Star-jet interaction model}
An alternative explanation of the fast variability can involve interactions of the ultrarelativisitic particles with compact objects entering the jet, e.g. stars or clouds \citep{bp97, ba10, br12, ar13, wy14, br15, bb15, de16, ba16}. 
In order to investigate if such a scenario can explain the fast variability observed in BL Lac in June 2015, we apply the model presented in \cite{ba16}. 
An emission region, i.e. the blob, filled with electrons is moving along the jet with a Lorentz factor $\Gamma$. 
It encounters and sweeps over a star that has entered the jet. 
The electrons propagating in the radiation field of the star produce $\gamma$ rays via inverse Compton process. 
Such a scenario can produce an orphan (VHE emission without any increase of optical flux), or nearly-orphan (sudden increase of VHE emission during a higher optical state without simultaneous increase of optical flux) VHE $\gamma$-ray flare.

The observed time scale of the emission is limited by two factors: the size of the region around the star, and the vertical dimensions of the blob, normally the latter one being the dominating one. 
In the model we assume that a star with a temperature $T=3\times10^4$\,K and radius $10^{12}$\,cm falls inside the jet. 
It is intercepted by a blob moving with $\Gamma=50$. 
The blob occupies a cross-section of the jet with a radius of $3.6\times10^{15}$\,cm.  These are similar to the values used for the blob in two zone modelling.
Its vertical size is described by a Gaussian profile with an RMS measured in the frame of the star of $7.2\times10^{13}$\,cm (i.e. in the frame of the jet both perpendicular and vertical sizes are comparable). 
The blob is filled with electrons injected with a power-law spectrum with an index of $2.35$ and the energy density of the electrons measured in the blob is $\sim 7.5\,\mathrm{erg\,cm^{-3}}$. 
The electrons interact with the stellar radiation field producing $\gamma$ rays. 
The $\gamma$ rays of high-enough energy crossing close to the star can be absorbed producing $e^+e^-$ pairs, which can in turn produce further $\gamma$ rays in an electomagnetic cascade. 
We calculate the spectra and light curves of the photons escaping at the typical observing angle of $\sim1/\Gamma$.
The model is then compared with the SED of MAGIC integrated in the 1.2\,h observations of the flare. 
Due to limited statistics and visibility window, it is not possible to use strictly simultaneous \textit{Fermi}-LAT data. 
Instead, we use a contemporaneous \textit{Fermi}-LAT data using the spectral index estimation from 12\,hrs around the MAGIC flare and absolute flux estimation for a 6-hr period. 
Longer integration time, combined with a vast variability of the source, can introduce an additional systematic uncertainty in the modelling; however the \textit{Fermi}-LAT and MAGIC spectra connect smoothly within the statistical uncertainties. 

In Fig.~\ref{fig:star} we show the comparison of the model with the $\gamma$-ray measurements. Within the statistical and systematic uncertainties the model can describe well the GeV-TeV spectrum and the light curve. 
The values of the Lorentz factor and the energy density in the blob, while large, are still less extreme than in the case of the PKS\,1222+21 flare \citep{al11, ba16}. 
We conclude that an interaction of a blob filled with relativistic electrons with the radiation field of a star is a viable explanation for the observed BL Lac flare.  

\section{Summary and Conclusions}

In this paper we have reported the detection of a fast VHE $\gamma$-ray
flare from BL Lac with halving time of $26\pm8$ minutes. The observations were
triggered by a high state of the source in the HE $\gamma$-ray (detected by
{\it Fermi}-LAT) and optical bands. We presented the MAGIC VHE $\gamma$-ray data together with multiwavelength data from 2015 May 1 (MJD 57143) to July 31 (MJD 57234).

The multiwavelength behaviour in this period is rather similar to
the behaviour that is typically observed in this source during high flux states;
the optical and $\gamma$-ray emission correlate, while X-ray
variability is less prominent. The optical polarization angle is
rotating $\sim$90$^\circ$ and we also detect a brightening in the
VLBA 43\,GHz core around the time of the $\gamma$-ray flare, locating
the active region there. The fast VHE $\gamma$-ray flare is
not accompanied by a significant brightening in $\gamma$-ray, X-ray or
optical bands; even though we have observations from the same day in all
of these bands. We note that, the X-ray observation is not strictly
simultaneous.

We also compare the multiwavelength behaviour to the two epochs around
the fast VHE $\gamma$-ray flares observed by VERITAS. We find that the
lower frequency patterns seem to repeat during the three fast VHE
$\gamma$-ray flares. As discussed in
\citet{icrc_veritas,2018arXiv180210113A}, the occurence of VHE
$\gamma$-ray flares during the activity in VLBA core is in agreement
with a model proposed by \citet{marscher14}. In that model the VLBA
core is a conical shock. Turbulent shells of plasma pass through that
conical shock and electrons are accelerated. As the turbulent shells
are small, they would naturally also explain the fast VHE $\gamma$-ray
flares. However, there is a significant observational bias involved in two of the three VHE $\gamma$-ray observations as they have been triggered by high
optical and $\gamma$-ray fluxes. We therefore considered three models to reproduce the SED in this paper.

\begin{figure}
\centering
\includegraphics[width=0.43\textwidth]{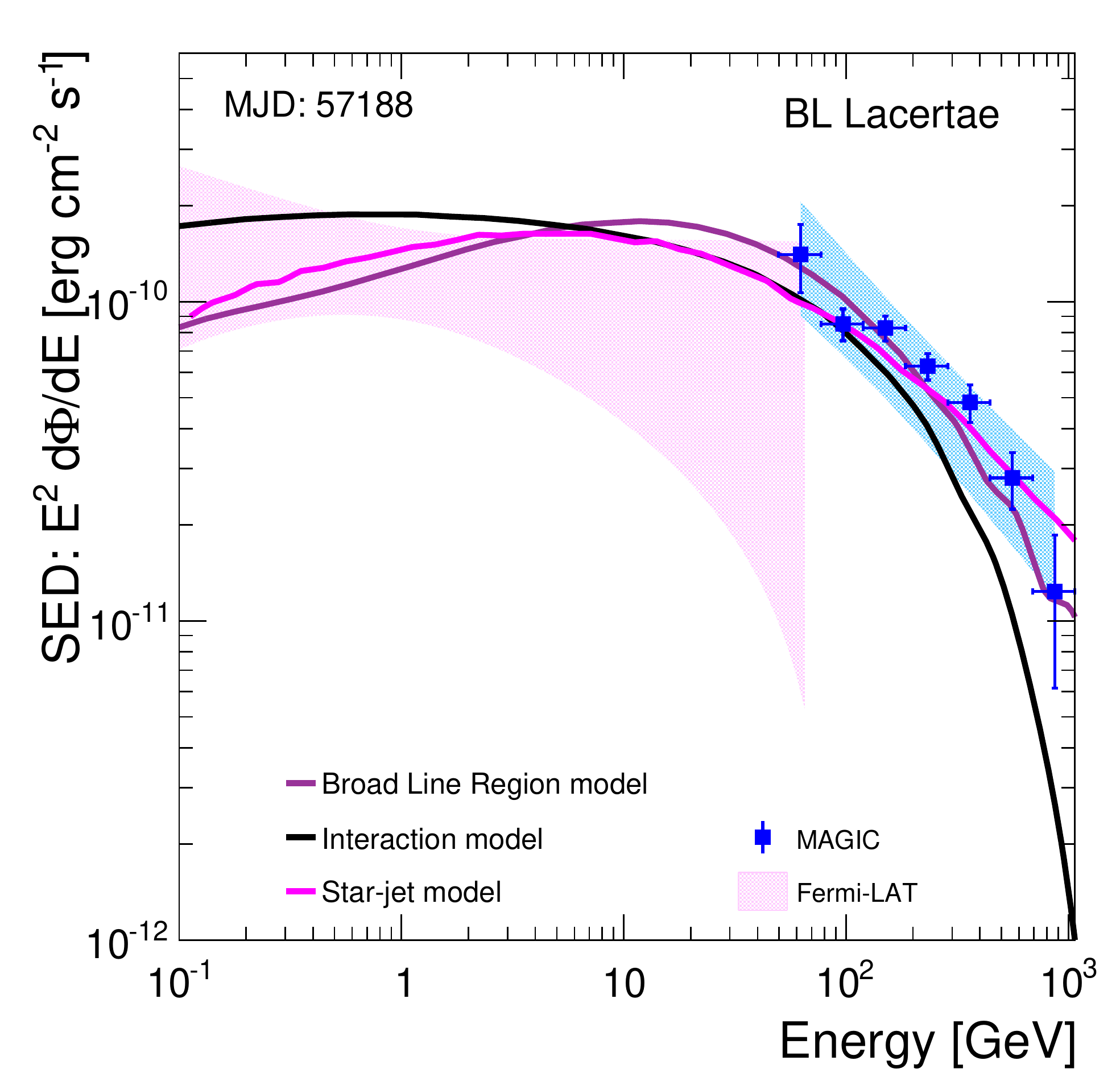}
\caption{Gamma-ray SED of MJD 57188 compared to the three models discussed in Section 4. The light blue band shows the systematic uncertainty of the MAGIC data.}
\label{fig:model_comparison}
\end{figure}

The models we consider (see Fig.~\ref{fig:model_comparison}), namely a fast blob
inside the BLR, a fast blob interacting with a larger
component and star-jet interaction, can all reproduce the observed
SED during the 2015 June 15 flare. All of
the models have some caveats. In the first model (small blob inside
BLR), there is large uncertainty about the parameters used for the
BLR, which is known to be weak in this source. In addition, there is no
spatial connection between the two emission regions, even if in all
observed cases of fast VHE $\gamma$-ray flares, we have seen activity
also in the VLBA 43\,GHz core. In the second model, where the emission
regions are co-spatial it is challenging to match the model with the highest energy MAGIC data without overproducing the flux in the {\it Fermi}-LAT band.
The model also requires us to use a lower magnetic field than what VLBA observations would indicate (assuming equipartition condition). This was also found for
PKS~1510-089 \citep{pks1510} when adopting a similar modelling. The star-jet
model has the same caveat as the small blob inside the BLR model; there is no connection between the generally increased flux levels in the other bands and the fast VHE $\gamma$-ray flare. Furthermore, fast VHE $\gamma$-ray flares seem to be rather common in BL Lac, as three have already been observed. It is rather
unlikely that all three would have been produced by star-jet
interaction \citep[][see discussion in]{ic310}.

In summary, we have tested three models to explain the fast variability of VHE $\gamma$-ray flux in BL Lac, but were not able to settle on preferred model. The interaction model is preferred as it matches the observed repeating multiwavelength patterns best, but in our simple blob-in-blob model it gives the worst description of the $\gamma$-ray band data.
 Further observations during VHE $\gamma$-ray flares are required with strictly simultaneous optical and X-ray high cadence data.
Repeating MWL patterns could play a key role in constraining the site and mechanism of fast $\gamma$-ray flares. This gives a strong motivation to have an intense long-term monitoring of BL Lac, regardless of its VHE $\gamma$-ray state.

\label{section:acknowledgements}
\begin{acknowledgements}
We would like to thank the Instituto de Astrof\'{\i}sica de Canarias
for the excellent working conditions at the Observatorio del Roque de
los Muchachos in La Palma. The financial support of the German BMBF
and MPG, the Italian INFN and INAF, the Swiss National Fund SNF, the
ERDF under the Spanish MINECO (FPA2015-69818-P, FPA2012-36668,
FPA2015-68378-P, FPA2015-69210-C6-2-R, FPA2015-69210-C6-4-R,
FPA2015-69210-C6-6-R, AYA2015-71042-P, AYA2016-76012-C3-1-P,
ESP2015-71662-C2-2-P, FPA2017‐90566‐REDC), the Indian Department of
Atomic Energy and the Japanese JSPS and MEXT is gratefully
acknowledged. This work was also supported by the Spanish Centro de
Excelencia ``Severo Ochoa'' SEV-2016-0588 and SEV-2015-0548, and
Unidad de Excelencia ``Mar\'{\i}a de Maeztu'' MDM-2014-0369, by the
Croatian Science Foundation (HrZZ) Project IP-2016-06-9782 and the
University of Rijeka Project 13.12.1.3.02, by the DFG Collaborative
Research Centers SFB823/C4 and SFB876/C3, the Polish National Research
Centre grant UMO-2016/22/M/ST9/00382 and by the Brazilian MCTIC, CNPq
and FAPERJ.  The work of the author M. Vazquez Acosta is financed with
grant RYC-2013-14660 of MINECO. F. D'Ammando is grateful for support
from the National Research Council of Science and Technology, Korea
(EU-16-001).

The \textit{Fermi} LAT Collaboration acknowledges generous ongoing support
from a number of agencies and institutes that have supported both the
development and the operation of the LAT as well as scientific data
analysis. These include the National Aeronautics and Space Administration
and the Department of Energy in the United States, the Commissariat \`a
l'Energie Atomique and the Centre National de la Recherche Scientifique /
Institut National de Physique Nucl\'eaire et de Physique des Particules in
France, the Agenzia Spaziale Italiana and the Istituto Nazionale di Fisica
Nucleare in Italy, the Ministry of Education, Culture, Sports, Science and
Technology (MEXT), High Energy Accelerator Research Organization (KEK) and
Japan Aerospace Exploration Agency (JAXA) in Japan, and the
K.~A.~Wallenberg Foundation, the Swedish Research Council and the Swedish
National Space Board in Sweden.

Additional support for science analysis during the operations phase is
gratefully acknowledged from the Istituto Nazionale di Astrofisica in
Italy and the Centre National d'\'Etudes Spatiales in France. This work
performed in part under DOE Contract DE-AC02-76SF00515.

Based on observations made with the Nordic Optical Telescope, operated by the Nordic Optical Telescope Scientific Association at the Observatorio del Roque de los Muchachos, La Palma, Spain, of the Instituto de Astrofisica de Canarias.

Acquisition and reduction of the MAPCAT data was supported in part by MINECO through grants AYA2010-14844, AYA2013-40825-P, and AYA2016-80889-P, and by the Regional Government of Andaluc\'ia through grant P09-FQM-4784. The MAPCAT observations were carried out at the German-Spanish Calar Alto Observatory, which is jointly operated by the Max-Plank-Institut f\"ur Astronomie and the Instituto de Astrof\'isica de Andaluc\'ia-CSIC.

The St.Petersburg University team acknowledges support from Russian
Science Foundation grant 17-12-01029.

This publication makes use of data obtained at the Mets\"ahovi Radio Observatory, operated by Aalto University, Finland.

This study makes use of 43 GHz VLBA data from the VLBA-BU Blazar Monitoring Program (VLBA-BU-BLAZAR;
http://www.bu.edu/blazars/VLBAproject.html), funded by NASA through the Fermi Guest Investigator Program. The VLBA is an instrument of the National Radio Astronomy Observatory. The National Radio Astronomy Observatory is a facility of the National Science Foundation operated by Associated Universities, Inc. The BU group acknowledges support from NASA Fermi GI program grant 80NSSC17K0694 and US National Science Foundation grant AST-1615796.

The OVRO 40-m monitoring program is supported in part by NASA grants NNX08AW31G, NNX11A043G and NNX14AQ89G, and NSF grants AST-0808050 and AST-1109911.
\end{acknowledgements}


\end{document}